\documentclass{aa}
\usepackage{graphicx}
\usepackage{txfonts}
\usepackage{longtable}
\usepackage{lscape}
\begin{document}
   \title{Comparing six evolutionary population synthesis models 
   through spectral synthesis on galaxies }

   \author{   X. Y. Chen \inst{1,2,3}\thanks{email: chenxy@bao.ac.cn},
             Y. C. Liang\inst{1,2}\thanks{email: ycliang@bao.ac.cn}, 
    F. Hammer\inst{4}, Ph. Prugniel\inst{5}, G. H. Zhong\inst{1,2,3}, 
    M. Rodrigues\inst{4}, Y. H. Zhao\inst{1,2}, H. Flores\inst{4}  }
    \institute{
    National Astronomical Observatories, Chinese Academy of Sciences, A20 Datun Road, 
    100012 Beijing, China
    \and
    Key Laboratory of Optical Astronomy, NAOC, Chinese Academy of Sciences
    \and
    Graduate School of the Chinese Academy of Sciences, 100049 Beijing, China 
    \and 
    GEPI, Observatoire de Paris, CNRS-UMR 8111, 5 place Jules Janssen, 92195 Meudon, France
    \and 
    Universit\'{e} Lyon 1, Villeurbanne, F-69622, France; CRAL, Observatoire de Lyon, St Genis Laval, F-69561, France; CNRS, UMR 5574
             }

   \date{Received ; accepted }


  \abstract
{}
{ We compare six popularly used evolutionary population synthesis (EPS) models 
through fitting the full optical spectra of six representative 
types of galaxies (star-forming and composite galaxies, Seyfert 2s, LINERs,
E+A and early-type galaxies), which are taken from the Sloan Digital Sky Survey (SDSS);
and we also explore the dependence of stellar population synthesis 
results on the main ingredients of the EPS models;
meanwhile we study whether there is an age sequence among these types of
galaxies.
}
{Throughout our paper, we use the simple stellar populations (SSPs) from each 
EPS model and the software STARLIGHT to do our fits. 
Firstly, to explore the dependence of stellar population synthesis on EPS models, 
we fix the age, metallicity, and initial mass function (IMF) 
to construct a standard SSP group. 
We then use the standard SSP group from each EPS model (BC03, CB07, Ma05, GALEV, GRASIL, 
and Vazdekis/Miles) to fit the spectra of 
star-forming and E+A galaxies. 
Secondly, we fix the IMF and change the selection of age and 
metallicity respectively to construct eight more SSP groups. 
Then we use these SSP groups to fit the spectra of star-forming and E+A galaxies 
to check the effect of age and metallicity on stellar populations. 
Finally, we also study the effect of stellar evolution track and stellar spectral library 
on our results. 
At the same time, the possible age sequence among these types of galaxies are suggested. 
}
{Using different EPS models the resulted numerical values of contributed light fractions change 
obviously, even though the dominant populations are consistent. 
The stellar population synthesis does depend on the selection of age and metallicity, 
while it does not depend on the stellar evolution track much. 
The importance of young populations decreases from star-forming, composite, Seyfert 2, 
LINER to early-type galaxies, and E+A galaxies lie between composite galaxies and Seyfert 2s 
in most cases.
}
{Different EPS models do derive different stellar populations, so that 
it is not reasonable to directly compare stellar populations estimated from different 
EPS models. 
To get reliable results, we should use the same EPS model for the compared samples.}
   \keywords{Galaxies: evolution -- Galaxies: Seyfert -- Galaxies: starburst -- 
   Galaxies: stellar content -- Stars: evolution }
\authorrunning {X. Y. Chen et al.}
\titlerunning {Comparing different EPS models through fitting spectra of galaxies}
\maketitle

\section{Introduction}
\label{introduction}
Stellar populations are fundamental characters in revealing the 
formation and evolution of galaxies. 
The formation of spiral galaxies is still hotly debated, and two main 
channels have been proposed, 
either initial collapse of gas at very high redshift in the frame of the 
tidal torque theory (Eggen, Lynden-Bell \& Sandage, 1962; White, 1984), 
or gaseous-rich mergers at intermediate to high redshifts (Hammer et al., 2005, 2007, 2009). 
These two channels for galaxy formation may provide distinct 
signatures from the analysis of the stellar populations, and it is 
relevant to test whether or not stellar population 
models can be used to test them (see for example Heavens et al., 2004; Panter et al, 2007).
While stars can not be resolved for a 
majority of galaxies, therefore many works have been generated on analyzing stellar 
populations through the integrated lights of the galaxies. 
Because the integrated lights hold information about age and metallicity distributions 
of their stellar populations and star formation histories.
This is the so-called stellar population synthesis on 
galaxies. Two main types of approaches have been developed: the 
empirical population synthesis (Faber 1972; Bica 1988; Boisson et al. 2000; 
Cid Fernandes et al. 2001) and the EPS (
Tinsley 1978; Bruzual 1983; Worthey 1994; Leitherer \& Heckman 1995; 
Maraston 1998; Vazdekis \& Arimoto 1999; Bruzual \& Charlot 2003;  Maraston 
2005; Cid Fernandes et al. 2005). 
In the empirical population synthesis approach, also known as 
$'stellar\,population\,synthesis\, with\, a\, data\, base'$, the observed 
spectrum of a galaxy is reproduced by a combination of spectra of 
individual stars or star clusters with different ages and metallicities from 
a  library.  The results following this approach do not consider the 
stellar evolution, and do not allow one to predict the past and future spectral 
appearance of galaxies.

The EPS approach 
uses the knowledge of stellar evolution to model the 
spectrophotometric properties of stellar populations, and has enjoyed 
more widespread use recently. In this approach, the main adjustable 
parameters are the stellar evolution tracks, the stellar spectral library, 
the IMF, the star formation history
(SFH), and the grids of ages and metallicities.  
EPS is a real
physical model, but it is restricted by the lacking of  comprehensive
stellar spectral library, accurate IMF and SFH, and  poor understanding of
some advanced phases of stellar evolution, such as  the blue stragglers
(BSs), the horizontal branch (HB) stars, and the 
thermally pulsating asymptotic giant branch (TP-AGB) stars. 

Up to now, several EPS models have been proposed and widely used 
in the stellar population studies on galaxies 
by analyzing their colors, spectra and multi-wavelength 
spectral energy distributions (SEDs), such as 
BC93 (Bruzual \& Charlot 1993), 
P\'{e}gase (Fioc \& Rocca-Volmerange 1997), 
GRASIL (Silva et al. 1998), 
GALAXEV (Bruzual \& Charlot 2003, BC03), 
CB07 (Charlot \& Bruzual 2009), 
SPEED (Jimenez et al. 2004), 
BaSTI (Pietrinferni et al. 2004, Cordier et al. 2007), 
Ma05 (Maraston 2005), 
Starburst 99 (V\'{a}zquez \& Leitherer 2005), 
Vazdekis/Miles (S\'{a}nchez-Bl\'{a}zquez et al. 2006; Vazdekis et al., in preparation), 
GALEV (Anders \& Alvensleben 2003; Kotulla et al. 2009), 
SPoT (Raimondo et al. 2005). 
  Some researches have used these EPS models to analyze the stellar populations 
  in galaxies and star clusters,
  and even compared them. The results are interesting, however, most of them
focus on analyzing the colors and multi-wavelength SEDs of the systems.
  
  Maraston et al. (2006) used two sets of EPS models to estimate 
 the star formation histories, ages, and masses 
of seven galaxies in the Hubble Ultra Deep Field by analyzing their
observed spitzer mid-IR (the rest-frame-UV) photometry data. 
One of the EPS models is Ma05, 
which includes the contribution of TP-AGB stars,
and another one is represented by BC03 (similar models are 
P\'{e}gase, Starburst 99 etc.). 
They concluded when they assumed a zero reddening, 
Ma05 gave better fits than BC03 for these distant 
passively evolving galaxies at $1.4<z<2.7$. While after dust was included in the fits, 
Ma05 performed no better than BC03.
Lee et al. (2007) reconstructed some composite grids by using 
BC03, Ma05, and SPoT in the color-color diagrams to estimate 
the average age and metallicity of spiral galaxies. They commented that 
the scatter among different models was large at ages $< 2Gyr$, and the dominant uncertainties 
arised from the treatment of different evolution phases (e.g. convective core overshoot, 
TP-AGB, helium abundance at higher metallicities). 
Longhetti \& Saracco (2008) found that the use of different models 
(BC03, Ma05, P\'{e}gase, and GRASIL) did not bring significant changes 
in the stellar mass estimates of early-type galaxies. Their samples were
10 massive early-type galaxies at $1<z<2$.
Muzzin et al. (2009) fit the UV to NIR SEDs of 34 K-selected galaxies at $z \sim 2.3$ 
(Kriek et al. 2006; 2007; 2008) by using BC03, Ma05, and CB07.
They concluded that there was no significantly better 
one among them, and the choice of the model produced more uncertainties 
than metallicities. 
Carter et al. (2009) reproduced the 
optical and near-infrared colours of 14 nearby elliptical and S0 galaxies
by using seven different EPS models 
(BC03; P\'{e}gase; Starburst 99; GALEV; SPEED; Ma05; BaSTI). Their results 
showed broad agreement on the ages and metallicities derived from different 
EPS models, although there were different deviations from the measured broad-band fluxes. 

Conroy et al. (2009a,b,c) carried out a series of works on studying the propagation of 
uncertainties in stellar population synthesis modeling. 
In their first work they explored the relevance of uncertainties in stellar evolution, 
IMF, and stellar metallicity distributions to the derived physical properties. 
They subsequently investigated some of the uncertainties associated with translating 
synthetic galaxies into observables, such as stellar evolution and dust. 
In their third work they especially performed the model calibration, comparison, and 
evaluation, and the models included their own flexible stellar population synthesis 
model (FSPS), BC03, and Ma05. 
They found that the FSPS and BC03 models were able to reproduce the optical and near-IR 
colors of E+A galaxies, while the Ma05 model performed poorly. 
They also pointed out significant differences between Ma05, BC03 and FSPS in terms of 
photometry of intermediate-age and sub-solar metallicities 
star clusters (i.e.star clusters in the MCs). 
Namely, their FSPS is better than BC03 and Ma05 for such cases.

Similarly, some other works have been applied to star clusters. 
Hempel et al. (2005) presented the results of optical and near-infrared photometry for 
globular cluster systems of two giant ellipticals. 
They compared the (V-H) and (V-I) colors from BC03 and Ma05, and  
found the color predictions for a given age did not differ significantly, except 
for metal-poor objects with ages $\leqslant 2 Gyr$. 
Pessev et al. (2008) adopted 54 star clusters to evaluate 
the performance of four EPS models 
(Vazdekis 1999, BC03, Ma05, GALEV) 
in the optical/near-infrared colour-colour space. They argued 
that each model had strong and weak points 
and there was no model that stood out in all aspects. 

There are also some works on analyzing the spectra of galaxies
or star clusters.
For example,
Panter et al. (2007) used MOPED to study 
the star formation history, the stellar mass function,
and the current stellar mass density in a large sample of SDSS DR3 galaxies. 
They also investigated the effects from the choices of 
the spectral resolution, sky lines, IMF, and EPS models 
(SPEED, P\'{e}gase, BC93, Ma05, BC03, CB07) on these properties. 
Their conclusion was that the main impacts on 
estimation of SFH were due to the EPS model, 
the calibration of the observed spectra and 
the choice of IMF. 
In 2008, they subsequently investigated the impact of model choices by
recovering the  metallicity history with the same six EPS models, and
suggested that older EPS models did not produce a clear results (Panter
et al. 2008).  
Cid Fernandes and his colleagues have worked
on spectral synthesis on the SDSS galaxies by using the BC03 model
and their STARLIGHT code. 
They also studied the properties of the galaxies, such as the dust extinction,
stellar mass, SFH etc. (Cid Fernandes et al. 2004, 2005, 2007;
Mateus et al. 2006; Asari et al. 2007, 2009; Stasi\'{n}ska et al. 2008). 
Koleva et al. (2008) compared different models, and 
found that the P\'{e}gase-HR (Le Borgne et al. 2004) 
and Vazdekis/Miles models were in precise agreement, 
while the BC03 model presented biases, which might be due to the poor metallicity coverage of 
STELIB library causing unreliable results at non-solar metallicities. 
In Koleva et al. (2009a), they used full-spectrum fitting to derive the radial profiles 
of the SSP-equivalent ages and metallicities for a sample of 16 dwarf elliptical galaxies
with their VLT spectra, and discussed the sensitivity to the 
population model and IMF.
 Recently, Cid Fernandes \& Gonzalez Delgado (2009) and
Gonzalez Delgado \& Cid Fernandes (2009) studied the ages, metallicities
and dust extinction of 27 star clusters from the Magellanic Clouds 
(from Leonardi \& Rose 2003) through stellar population synthesis. 
They adopted STARLIGHT to fit their integrated optical spectra in the blue-near-UV range
(3650-4600\AA). They further compared the combinations of model and 
spectral library.

Although these impressive progresses in stellar population analysis of galaxies
and in comparisons among different EPS models,
we can easily notice that most of them trade on colors and 
multiwavelength SEDs of galaxies. 
There are no much work through fitting the full spectra of 
galaxies, i.e. the detailed fittings on their stellar absorptions and continua.
Although some works do fit spectra to study the properties
of galaxies, there are no much efforts on comparing the different EPS models 
through fitting spectra on stellar population analysis.
Therefore, in this work, we will compare six popularly used EPS models
carefully through fitting the good quality optical spectra of a 
representative sample of galaxies
taken from SDSS. 

The powerful SDSS has provided good quality
spectra of hundred thousands galaxies 
and many other types of astronomical objects 
(such as AGNs) with middle resolution (3\AA, $R$=2000) 
(Tremonti et al. 2004; Kauffmann et al.  2003; Brinchman et al. 2004 etc.). 
These are good samples for reliable studies on stellar populations 
of galaxies from full-spectrum fitting on continua and stellar absorption. 
It also provides a good way to compare the different EPS models.
 
In this work, 
we will compare six different EPS models 
(BC03, CB07, Ma05, GALEV, GRASIL and Vazdekis/Miles),
by doing spectral synthesis analysis on six representative types of
galaxies from SDSS:
419 star-forming, 326 composite galaxies, 35 Seyfert 2s, 69 LINERs,
502 E+A galaxies, and 754 early-type galaxies.
Hence, we check the dependences of the main ingredients of EPS models,
i.e. the ages, metallicities, and stellar evolution tracks etc., on the stellar population
analysis results.
 
We will try to answer these questions: 
Whether these different EPS models (with same ingredients)
could provide same/similar stellar population on one galaxy?  
If there is difference,
what is the main ingredient 
dominates it?
What about the stellar populations of these representative galaxies?
Is there any age sequence among them? 
The latter two questions are also very interesting and important, although 
our main aim is to check some aspects of the consistency between models 
by using a single program.

This paper is organized as follows. 
In Sect.~\ref{sec.models}, 
the detailed illustration about the six EPS models and their
main ingredients are 
performed. The sample selections are given in 
Sect.~\ref{sec.sample}. 
In Sect.~\ref{sec.results}, we present the methods and the spectral synthesis results 
from different EPS models, 
and then compare these models. 
The dependences of ages and metallicities on stellar populations
of galaxies are also checked.
In Sect.~\ref{sec.other}, we discuss the dependences of stellar populations
of galaxies on stellar evolution tracks, and also 
present the possible
age sequence among different types of galaxies.
Summary and conclusions are given in Sect.~\ref{sec.summary}.
Throughout this paper we assume the following cosmological 
parameters: $\Omega_{M}=0.3$, $\Omega_{\Lambda}=0.7$, and $H_{0}$=70 km\,s$^{-1}$\,Mpc$^{-1}$.

\section{EPS models}
\label{sec.models}

In this section, 
we will present the six EPS models which we will use and compare
in stellar population analysis of galaxies.
They are BC03, CB07, Ma05, GALEV, GRASIL and Vazdekis/Miles. 
The main ingredients of them contain 
the stellar evolution tracks, the stellar library, 
the grids of ages and metallicities,
the IMF and the SFH.

\subsection{The main ingredients of EPS model}

One of the important ingredients in EPS model is the stellar evolution track,
which records the  evolution of stars of any mass from the zero-age main
sequence to later evolutionary stages.  The tracks with wide range of mass
and time are necessary for EPS, and many groups have  produced public
stellar evolution models:  Padova\,1994 (Pa\,94, Alongi et al. 1993; Bressan
et al. 1993; Fagotto et al. 1994 a,b;  Girardi et al. 1996) includes all
phases of stellar evolution from zero-age main sequence to  the beginning of
TP-AGB (for low- and intermediate- mass stars) and core-carbon ignition 
(for massive stars);  Padova\,1999 (Pa\,99) extends earlier models by
the inclusion of the TP-AGB phase for  stars in the mass range $2 M_{\odot}
\leq m \leq 7 M_{\odot} $ in accordance with the fuel  consumption theorem
(Schulz 2002);  Padova\,2000 (Pa\,00, Girardi et al. 2000); Geneva (Schaller
et al. 1992);  Cassisi (Cassisi, Castellani \& Castellani 1997; Cassisi,
Degl'Innocenti \& Salaris 1997  and Cassisi et al. 2000); and Marigo \&
Girardi (2007), which is a new synthetic models of  the TP-AGB evolution. 

The stellar spectral library is another important ingredient in EPS model. 
An ideal stellar  library should provide complete coverage of the HR diagram,
accurate atmospheric  parameters ($T_{eff}$, surface gravities log $g$,
metallicities $Z$ etc.), good  wavelength coverage, good spectral resolution
and good calibration.  Bruzual (2005) and Coelho (2009) reviewed some
characteristics of different stellar spectral  libraries.   
There are two types of stellar spectral libraries: empirical library and
theoretical  library. The empirical library (eg.  STELIB, Le Borgne et al.
2003; UVES POP, Valdes et al. 2004; Indo-US, Jehin et al. 2005;  MILES,
S\'{a}nchez-Bl\'{a}zquez et al. 2006; ELODIE, Prugniel et al. 2007)  is
based on observations of real stars, so that it is more  reliable. However,
it is limited by the quality of the observations, thus the coverage  of
parameters are biased towards the typical stellar population targeted by the
observations.  While the theoretical spectral library (eg. Kurucz 1992;
BaSeL 1.0, Lejeune et al. 1997, 1998)  is based on model atmospheres, so
that its coverage of parameters can be at will, but limited  by our
knowledge of the physics of stellar atmospheres. 

The third important ingredient in EPS model is the IMF, i.e. 
the initial distribution of the stars along the main sequence. 
Pioneered by Salpeter (1955), many types of IMF emerged 
(Kennicutt 1983; Scalo 1998; Kroupa 2001; Chabrier 2003; etc.). 
The general form of IMF is $\phi(M) \varpropto M^{-(1+\alpha)}$, and 
the logarithmic slope and the upper and lower limits in IMF are particular relevant to EPS. 

 With the above three main ingredients 
(stellar evolution track, stellar spectral library, and the 
IMF), we can construct SSPs. 
Then the time-dependent SSPs are convolved with an arbitrary SFH to synthesis 
the spectrum or colors of a galaxy (Kennicutt 1998). 
Throughout this work, we apply a SFH of instantaneous burst 
to our analysis i. e. $\psi(t)=1M_{\odot}\,\delta(t)$.

\begin{table*}
\caption{Comparisons among different EPS models. 
First, the stellar spectral library, delivering to the spectral resolution, the range and 
number of wavelength, are fixed in our work; 
Second, in this table, we list the full range of age and metallicity, and we will select 
some of them to construct the groups of SSPs for our fit (Table~\ref{agez.tab}); 
Third, there are several choices of IMFs and stellar evolution tracks in some models, and we 
only list the one we frequently used. 
}
\centering
\label{para.tab}
\begin{tabular}{l|l|l|l|l|l|l}
\hline
models& BC03&CB07 & Ma05 & GALEV&  GRASIL & Vazdekis/Miles\\
\hline
library&STELIB/BaSeL3.1&STELIB/Kurucz92& BaSeL2.0& BaSeL2.0&Kurucz1992& Miles2006\\
resolution$(\AA)^{a}$& 3& 3& 20 & 20 & 20& 2.3\\
wavelength ($\AA$) &91-1.6x10$^{6}$& 91-3.6x10$^{8}$& 91-1.6x10$^{6}$& 91-1.6x10$^{6}$&91-1.2x10$^{7}$ & 3540-7410\\
N$_{\lambda}$& 6900 & 6917 & 1221 & 1221 & 1264& 4300\\
age(Gyr)(Number)&0-20(221)&0-20(221)&10$^{-6}$-15(67)&4x10$^{-3}$-16(4000)&10$^{-4}$-20(55) & 0.06-18(50)\\
Z(Number)&0.0001-0.05(6)&0.0001-0.05(6)& 0.0001-0.07(6)& 0.0004-0.05(5)&0.0001-0.1(7)& 0.0001-0.03(7)\\
IMF&Salpeter &Salpeter&Salpeter& Salpeter &Salpeter&Salpeter\\
track&Pa\,94&Pa\,94+Marigo\,07& Cassisi+Geneva& Pa\,99&Pa$^{b}$ & Pa\,00\\
\hline
\end{tabular}
\begin{list}{}{}
\item{Notes:
$^a$: resolution in visual regions.
$^b$: refer to Bertelli et al. 1994.}
\end{list}
\end{table*}

\subsection{BC03}
\label{sec.bc03}
A popular library of evolutionary stellar population synthesis models,
GALAXEV, is  computed by using the isochrone synthesis code BC03 (Bruzual \&
Charlot 2003).  The SSP models given by BC03 span a large range of
wavelength ($91\AA \sim 160\mu\,m$, N=6900),  age ($0 \sim 20\,Gyr$,
N=221), and metallicity Z ($0.0001 \sim 0.05$, N=6).  These models mainly
use the STELIB/BaSeL3.1 libraries (Le Borgne et al. 2003; Lejeune et al. 
1997,1998; Westera et al. 2002 and references therein) with  resolutions of
$3\AA$ (FWHM) from 3200 $\AA$ to 9500 $\AA$ and $20\AA$ (FWHM) elsewhere. 
Besides that, BC03 also provides models rely on STELIB/Pickles libraries
(Pickles 1998).  Moreover, there are 2 IMFs (Chabrier, Salpeter) and 3
stellar evolution track (Pa\,94, Pa\,00, and Geneva) we can choose. 
These information are given in Table~\ref{para.tab}.

\subsection{CB07}
\label{sec.cb07}
CB07 is a new version of BC03, which includes the new stellar evolution
prescription of  Marigo \& Girardi (2007) for the TP-AGB 
evolution of low- and intermediate-mass
stars.  It is given by Charlot \& Bruzual (2009),
and we obtained the model by private communications.  The SSP
models given by the current version of CB07 span a range of wavelength 
($91\AA \sim 36000\mu\,m$, N=6917),  age ($0 \sim 20\,Gyr$, N=221), and
Z($0.0001 \sim 0.05$, N=6).  Moreover, there are two choices of IMF:
Chabrier and Salpeter. 
These information are given in Table~\ref{para.tab}.

\subsection{Ma05}
\label{sec.ma05}
The original version of Ma05 is Ma\,98, 
which is an EPS model based on fuel consumption theorem (quite different from other models). 
A grid of SSP models with $Z=Z_{\odot}$ and 
an age range of $30 Myr \sim 15 Gyr$ was constructed. 
Ma\,05 spans a wider range of stellar population 
parameters: 6 metallicities ($Z=0.0001\sim 0.07$), 
67 ages ($10^{3}yr \sim 15\,Gyr$), 2 IMFs (Salpeter, Kroupa), and 3 horizontal branch (HB) 
morphologies (red, intermediate or blue; see details in Maraston 2005). 
However, not each metallicity match all 67 ages:
$Z=0.0001$ associated to 16 ages ($1 \sim 15 Gyr$) with Cassisi tracks; 
$Z=0.07$ associated to 16 ages ($1 \sim 15 Gyr$) with Pa\,00 tracks; 
while the rest 4 metallicities associated to full 67 ages with Cassisi + Geneva tracks. 
Moreover, the Cassisi tracks are without considering overshooting, 
while the Padova tracks include the effects of overshooting. The stellar spectra were taken 
from BaSel library (Lejeune et al. 1998), with low spectral resolution, i.e. $5-10\AA$ up to 
the visual region, $20-100\AA$ in the near-IR (wavelength ranges from $91\AA$ to $160\mu\,m$, 
and $N_{\lambda}=1221$). 
These information are given in Table~\ref{para.tab}.

\subsection{GALEV}
\label{sec.galev}
GALEV (GALaxy EVolution) evolutionary synthesis models describe the spectral and chemical 
evolution of galaxies over cosmological timescales, i.e. from the beginning of star formation 
to the present (Kotulla et al. 2009). This code considers both the chemical evolution of the 
gas and the spectral evolution of the stellar component, allowing for what they call a 
chemically consistent treatment. 
Thus some SSPs in this model present emission lines in their spectra. 
Additionally the GALEV evolutionary synthesis models are 
interactively available at web-interface\footnote{http://www.galev.org}.
The SSP models provided by this code cover 5 metallicities 
($0.02 \leqslant Z/Z_{\odot} \leqslant 2.5$), and 4000 ages ($4 \times 10^{6} yr \sim 16 Gyr$) 
in time-steps of $4 Myr$. They are based on the spectra from Lejeune et al. 
1997,1998 (BaSeL 2.0), on 3 IMFs 
(Salpeter, Scalo, and Kroupa), and on the theoretical isochrones from the Pa\,99 
and the Geneva tracks. 
The lower limit of stellar mass for IMF is always $0.1 M_{\odot}$, while the upper mass limits are as follows: 
for Padova, it is set by isochrones (about $50 M_{\odot}$ for super-solar metallicity 
and about $70 M_{\odot}$ for the rest); for Geneva, it is always $120 M_{\odot}$.
The range of wavelength is $90\AA \sim 
160 \mu\,m$ with resolution of $20 \AA$ in the UV-optical and $50 - 100 \AA$ in the NIR ranges 
(Schulz et al. 2002; Anders \& Alvensleben 2003).
These information are given in Table~\ref{para.tab}.

\subsection{GRASIL}
\label{sec.grasil}
GRASIL (GRAphite and SILicate) is a population synthesis code, which takes into full account 
the effects of dusty interstellar medium in galaxy spectra (Silva et al. 1998). It is a multi-wavelength model 
for the combination of stellar population and dust, which absorbs and scatters optical and 
UV photons and emits in the IR-submm region. It is particularly suited to the study of the IR 
properties of dusty galaxies.
Additionally, GRASIL can be very conveniently run using the web interface GALSYNTH \footnote{
http://galsynth.oapd.inaf.it/galsynth/index.php}.
The SSP models given by this code spans a large range of metallicity 
($Z=0.0001\sim 0.1$, N=7), and age ($10^{5}yr \sim 20Gyr$, N=55). 
Moreover the models are based on the Kurucz (1992) stellar atmosphere model, 
on 4 IMFs (Salpeter, Kennicutt 1983, Miller \& Scalo and Scalo) with mass range $0.15\sim120M_{\odot}$, 
and on the Padova 
tracks (Bertelli et al. 1994) adding the effects of dusty envelopes around AGB stars (see details in Silva's PhD thesis). 
Besides that, the spectra range from $91\AA$ to $1200\mu\,m$ 
with resolution of 20 $\AA$. 
These information are given in Table~\ref{para.tab}.

\subsection{Vazdekis/Miles}
\label{sec.vazd}
The EPS model explored by Vazdekis et al. started from 
1996, experienced several generations, and predicted for 
studying old and intermediate aged stellar populations. 
The original version V96 (Vazdekis et al. 1996) used the Lick polynomial fitting functions, which were based 
on the Lick/IDS stellar library, however, the Lick stellar library has not been flux 
calibrated (V96). 
Then, in 1999 Vazdekis gave an extended version of V96 (Vazdekis 1999, V99), 
which provided flux-calibrated spectra. This version (V99) predicted 
SEDs for SSPs in two reduced spectral regions in the optical wavelength range 
($3855 \sim 4476 \AA, 4795 \sim 5465 \AA$) at resolution $\sim 1.8 \AA$ (FWHM), 
with a range of Z 
($-0.7\, \leqslant\,log(Z/Z_{\odot})\,\leqslant\,+0.2$) and ages (1 to 17 Gyr). The input 
stellar database is the empirical stellar library of Jones (1999). There are 4 types of IMFs 
provided by this version: Unimodal, Bimodal, Kroupa universal, and Kroupa revised (Vazdekis 
et al. 2003 and references therein). 
Further, a revised version of the model of V96 and V99 was explored (Vazdekis et al. 2003), 
which replaced the database of Jones (1999) by Cenarro (2001), 
and replaced the isochrones of Bertelli (1994) by Pa\,00. 
This model predicted both the strength of the Ca\,II triplet feature and SEDs in the range 
$8449 \sim 8952 \AA$ at resolution $1.5 \AA$ (FWHM), with metallicities 
$-1.7\,<\,[Fe/H]\,<\,+0.2$, ages $0.1\,\sim\,18\,Gyr$, and 4 types of IMFs (same as V99).
In this code, we adopted the latest version based on a new empirical stellar spectral library 
MILES (S\'{a}nchez-Bl\'{a}zquez et al. 2006). The model spans the range of 
wavelengths $3540 \sim 7410 \AA$ at resolution $\sim 2.3 \AA$ (FWHM), 7 metallicities 
$-2.32\,\leqslant\,log(Z/Z_{\odot})\,\leqslant\,+0.22$ with Pa\,00 tracks, 
and 50 ages $0.063\sim 17.78\,Gyr$ (Vazdekis et al., in preparation).
These information are given in Table~\ref{para.tab}.

We summarize the related parameters of all the six EPS models in Table~\ref{para.tab},  
which are the most frequent choices in stellar population analysis and
used in this work.

\section{Sample}
\label{sec.sample}
We selected six representative types of galaxies as our working samples.
They are
849 emission-line galaxies including 419 star-forming galaxies, 326 composite
galaxies, 35 Seyfert 2s and 69 LINERs (taken from Chen et al. 2009);
502 E+A galaxies (from Goto 2007); 
and 754 early-type galaxies (from Hao et al, 2006).
We downloaded all the 1D spectra of them from the SDSS,
which have been sky-subtracted, then the telluric absorption bands were 
removed before the wavelength and spectrophotometric were
calibrated (Stoughton et al. 2002). 
We corrected the foreground Galactic extinction using the reddening maps of 
Schlegel et al. (1998) and then shifted the spectra to the rest frame. 
In order to view the global properties
of the galaxies in each of the sub-groups, we combined all the spectra 
in each of the sub-groups using the task SCOMBINE in IRAF. 
That is, we combined spectra by interpolating them to a common dispersion sampling, 
and computing the median of the pixels.
In this way we can also improve well the S/N of the spectra ($26\sim 56$ in 
$4730\AA \sim4780\AA$). 
Then we work on these six combined spectra. 

\subsection{Emission-line galaxies}
\label{sec.elg}

The emission-line galaxies were taken from Chen et al. (2009) directly. 
This sample was selected from the main galaxy sample of the
Sloan Digital Sky  Survey (SDSS) DR4 database and then was cross-correlated
with the IRAS Point Source  Catalog (PSCz) ($5 \arcsec$ matching radius). We
then selected  high signal-to-noise ratio (S/N) targets according to the
criteria of  S/N greater than $5\sigma$ for $H\,\beta, H\,\alpha,
[NII]\lambda6583 $, and S/N greater than  $3\sigma$ for
$[OIII]\lambda5007$.  Further the objects with spectra not located at the
center of galaxies  , and those with problematic mask spectra were all
removed from our sample (see details in  Chen et al. 2009). Finally, 849
objects left in our emission-line galaxies' sub-sample. The observed-frame
spectral wavelength range is 3800-9200 \AA, and the resolution is 3 \AA (
FWHM). 

These emission-line galaxies were further divided into 4 groups: 419 star-forming
galaxies, 326 composite galaxies, 35 Seyfert 2s and 69 LINERs, according to the
emission-line diagnostic diagram (see figure $2$ in Chen et al. 2009; 
Baldwin et al. 1981, BPT; Shuder et al.
1981; Veilleux \& Osterbrock 1987;  Kewley et al. 2001; Kauffmann et al. 2003). 
Composite galaxies here refer to objects whose spectra contain significant contributions 
from both AGN and star formation (Brinchmann et al. 2004; Kewley et al. 2006; 
Chen et al. 2009). 
Most of them have redshift $z<0.1$ due to the limit of IRAS observations. 

\subsection{E+A galaxies}
\label{sec.kpa}
E+A galaxies are so called because their spectra look like a superposition 
of elliptical galaxies and of A-type stars.  On the one hand, the features
of elliptical are not in point of morphology (although most of  them are
elliptical) but in point of non-detection of ongoing star formation
indications, i.e.  these galaxies' spectra do not show significant emission
lines.  On the other hand, the features of A-type stars are reflected in
strong Balmer absorption  lines. Therefore E+A galaxies have been explained
as post-starburst galaxies, that is, a galaxy has experienced
starburst recently, but truncated it suddenly. Some works have been
performed on this type of galaxies ( Falkenberg et al. 2009a, b; Huang \& Gu
2009; Pracy et al. 2009, and references therein). 

We selected our sample of E+A galaxies from the catalogue of Goto 2007,
which was based on the SDSS DR5, and was an extended sample from Goto 2005 on DR2.  
There were 564 E+A galaxies in this
catalogue, and the selection algorithm is summarized as follows:  they
only used objects classified as galaxies, spectroscopically classified not
to be a star,  and with spectroscopic S/N$ > 10$ per pixel to remove the
pollution from nearby stars and  star-forming regions;  then they selected
E+A galaxies as those satisfied the criteria of  H$\delta$ equivalent width
(EW)$ > 4 \AA$\footnote{Absorption lines have a positive EW.},  [OII] EW$ >
-2.5 \AA$, H$\alpha$ EW $> -3.0 \AA$;  moreover, they excluded galaxies at
$0.35 < z < 0.37$ due to the sky feature at $5577 \AA$.  From this
catalogue, we further discarded 62 E+A galaxies as they have problematic
mask  spectra. Thus our final sample of E+A galaxies contains 502 objects. 

\subsection{Early type galaxies}
\label{sec.ear}
Early-type galaxy has been the most studied objects currently, because of 
its relatively little dust extinction, gaseous interstellar medium, and little recent 
star formation (Vazdekis et al. 1996; Cid Fernandes et al. 2009, and references therein). 

Our sample of early-type galaxies was selected from Hao et al. (2006), which was based on the 
SDSS DR4 photometric catalogue: 
they selected objects with redshifts smaller than 0.05, with the velocity dispersions cover a 
range of 200 km s$^{-1}$ to 420 km s$^{-1}$, and not saturated or located at the edge of the 
corrected frame; they further excluded some objects which were unavailable from the SDSS DR4 
Data Archive Server (DAS), and those with visible dust lanes. 
Then they visually examined all images to confirm the target objects were E/S0 galaxies and not 
contaminated by companion galaxies or bright stars. 
Moreover, we discarded 93 objects with problematic mask spectra, so that our final sample 
of early-type galaxies contains 754 objects. 

\begin{table*}
\caption{Selections of ages and metallicities for SSPs from each model, which 
we will use below. BC03 and CB07 are shown in the same line. See more details in Sect.~\ref
{sec.ssp}}
\centering
\label{agez.tab}
\begin{tabular}{l|cccc|cccc|cccc|ccc}
\hline
models   & \multicolumn{12}{c|}{ages(Gyr)}& \multicolumn{3}{|c}{metallicities ($Z_{\odot}$)}\\
\hline
 & \multicolumn{4}{c|}{Young ($<0.2Gyr$)} &\multicolumn{4}{c|}{ Intermediate 
 ($0.2Gyr\sim 2Gyr$)} & 
\multicolumn{4}{c|}{Old ($>2Gyr$)} &   &  & \\
\cline{2-13}
& Y1 & Y2 & Y3 & Y4 & I1& I2& I3& I4& O1& O2& O3& O4& Z1& Z2& Z3\\
\hline
BC03,CB07 & 0.004 & 0.010 & 0.064 & 0.102&  0.286& 0.509& 0.905& 1.434&  3.00& 6.00 & 10.00 & 13.00&  0.2& 1.0 & 2.5 \\
Ma05 & 0.004 & 0.010 & 0.065 & 0.100&  0.300& 0.500& 0.900& 1.500&  3.00& 6.00 & 10.00 & 13.00& 0.5&  1.0 & 2.0 \\
GALEV & 0.004 & 0.012 & 0.064 & 0.100&  0.280& 0.500& 0.900& 1.432&  3.00& 6.00 & 10.00 & 13.00&  0.2& 1.0 & 2.5 \\
GRASIL & 0.004 & 0.010 & 0.060 & 0.100&  0.300& 0.500& 0.900& 1.500&  3.00& 6.00 & 10.00 & 13.00& 0.2&  1.0 & 2.5 \\
Vazdekis/Miles &  &  & 0.063 & 0.100&  0.280& 0.500& 0.890& 1.410&  3.16& 6.31 & 10.00 & 12.59&  0.2& 1.0 & 1.5 \\
\hline
\end{tabular}
\end{table*}


\section{Comparisons among the six EPS models based on spectral synthesis}
\label{sec.results}

\subsection{Spectral synthesis method}
\label{sec.method}
We fit the spectral absorptions and continua of the sample galaxies to study their stellar 
populations by using the software STARLIGHT\footnote{http://www.starlight.ufsc.br} 
(Cid Fernandes et al. 2005, 2007; Mateus et al. 2006; Asari et al. 2007; Chen et al. 2009). 
It is a program to fit an observed spectrum $O_{\lambda}$ with a model $M_{\lambda}$ that adds 
up to $N_{\ast}$ SSPs with different ages and metallicities from different stellar population 
synthesis models. A Gaussian distribution centered at velocity $v_{\ast}$ and broadened by 
$\sigma_{\ast}$ models the line-of-sight stellar motions. The fit is carried out with the 
Metropolis scheme (Cid Fernandes et al. 2001), which searches for the minimum 
$\chi^{2}=\Sigma_{\lambda}[(O_{\lambda}-M_{\lambda})\omega_{\lambda}]^{2}$, where the 
reciprocal of weight $\omega_{\lambda}^{-1}$ is the error in $O_{\lambda}$ except for masked regions. Pixels that 
are more than $3\sigma$ away from the rms $O_{\lambda}-M_{\lambda}$ are given zero weight by 
the parameter $'clip'$. 
 The STARLIGHT group has carefully checked the reliability of 
 this software by analyzing the ``stellar populations" 
 of fake galaxies made by known SSPs (see figure4 in Cid Fernandes et al. 2005,
 and figure1 in Cid Fernandes et al. 2004).
 
Throughout our fit, we used the reddening law of 
Calzetti et al. (1994, CAL hereafter). 
The fitted wavelength range was from 3700 to 7400 \AA.
A power law stands for non-stellar component (Koski et 
al. 1978) $F_{\nu}\varpropto\nu^{-1.5}$ was added when we fit the spectra of Seyfert 2s and 
LINERs. The mask regions were the same as Chen et al. (2009), 
 which include
the strong emission-lines and the bad pixels. 
Additionally, we do not consider different weights in different regions of the spectra, 
since we have tested this and found that it did not affect our spectral synthesis results. 

In the outputs of STARLIGHT, one of 
the most important parameters to present stellar population is the population vector $\vec{x}$. 
The component $x_{j} (j=1,...,N_{\ast})$ represents the fractional contribution of the SSP with 
age $t_{j}$ and metallicity $Z_{j}$ to the model flux at the normalization wavelength 
$\lambda_{0}=4020\AA$. Equivalently, another important parameter, the mass fraction $\mu_{j}$, 
has the similar meaning.

\subsection{SSP selections}
\label{sec.ssp}
As mentioned in Sect.~\ref{sec.models}, we select 12 (or 10) representative
ages and 3 metallicities from 
each EPS model to construct different SSP groups to do our fits, and we list the details in 
Table~\ref{agez.tab}. 
We arrange the ages of SSPs into 3 bins: young populations with age $<0.2Gyr$, 
intermediate-age populations with age between $0.2Gyr$ and $2 Gyr$, 
and old ones with age $>2Gyr$. 
We note that these criteria are different from Chen et al. (2009), which followed the 
work of Kong et al. (2003) (i.e. young with age $<0.58Gyr$, old with age $>10Gyr$, and 
intermediate with ages between these two). 
Since the width of Balmer absorption line can indicate the young population with age $<1Gyr$, 
while the ones older than $1.5-2Gyr$ will be dominated by emission lines. 
Generally from SDSS spectra it is uneasy to disentangle $3-5 Gyr$ population from $10 Gyr$ 
populations, and Mathis et al. (2006) also pointed out that the signatures of 
intermediate-age stars ($0.5 \sim 4 Gyr$) are masked by those of younger and older stars, 
so we adopt the new criteria here. 

 The logic of our work is in five steps as follows:
 
\begin{enumerate}
\item First, we fix some main ingredients for the EPS models, i.e., 
the stellar library, the stellar evolution track and the IMF (Salpeter) 
are selected according to Table~\ref{para.tab}.
An instantaneous burst SFH is always applied.

\item Then we select the first SSP group including 6 SSPs at 6 different ages and 
at solar metallicity (see Table~\ref{agez.tab})
from each EPS model. It is called group {\bf No.(1)} and  
as our ``standard case" 
(see Table~\ref{groups.tab}): 
two young SSPs (Y1=0.004, Y2$\sim$0.01 Gyr, except for Vazdekis/Miles, in which 
the youngest ages Y3$\sim$0.06 and Y4$\sim$0.10 Gyr are selected),
two intermediate SSPs (I2$\sim$0.5, I3$\sim$0.9 Gyr) and 
two old SSPs (O3$\sim$10, O4$\sim$13 Gyr). 
This ``standard case" will be used to analyze the stellar populations of star-forming 
and E+A galaxies in Sect.~\ref{sec.compare}.
 
\item Next, we construct another three SSP groups 
by adding/removing some young, intermediate and old SSPs 
at solar metallicity. Then we can check the age dependences. 
The corresponding three groups are: 

{\bf No.(2):} 3 SSPs with ages of Y1 (Y3 for Vazdekis/Miles), I2, and O3 at $Z_{\odot}$ (Z2);

{\bf No.(3):} 9 (or 8) SSPs with ages of Y1, Y2, Y3, I2, I3, I4, O2, O3, O4 (only Y3 and Y4
for the young populations in Vazdekis/Miles); 

{\bf No.(4):}  12 (or 10) SSPs with ages of  Y1, Y2, Y3, Y4, I1, I2, I3, I4, O1, O2, O3, O4
 (only Y3 and Y4 for the young populations in Vazdekis/Miles);

\item After that, we construct another three SSP groups
by adding SSPs with 
sub-solar metallicity or/and super-solar metallicity 
based on group No.(1). 
Then we can check the metallicity dependences. 
These three groups are:

{\bf No.(5):}  12 SSPs with 6 ages in common with the group No.(1) but at two metallicities: 
  Z1 (0.2 or 0.5 $Z_{\odot}$) and Z2 ($Z_{\odot}$);

{\bf No.(6):}  12 SSPs with 6 ages as in group No.(1) but at two metallicities: 
  Z2 ($Z_{\odot}$) and Z3 (2.0, 2.5 or 1.5 $Z_{\odot}$);

{\bf No.(7):}  18 SSPs with 6 ages as in group No.(1) but at three metallicities: 
  Z1, Z2 and Z3.
  
\item Finally, we fix the ages and only change the metallicity of group No.(1) to 
build the last two SSP groups. Such that we can check the age-metallicity 
degeneracy in simple way. 

{\bf No.(8):} 6 SSPs with 6 ages as in group No.(1) but at sub-solar metallicity (Z1); 

{\bf No.(9):} 6 SSPs with 6 ages as in group No.(1) but at super-solar metallicity (Z3).

Table~\ref{groups.tab} list the details about the 9 SSP groups No.(1)-(9).

\end{enumerate}

\begin{table*}
\caption{Seven SSP groups used in our stellar population analysis. 
Please see Table~\ref{agez.tab} for the meanings of the symbols.
``VM" refers to the model Vazdekis/Miles. Group No.(1) is our ``standard case", 
and is the basis of other groups.}
\centering
\label{groups.tab}
\begin{tabular}{l|l|l|l|l}
\hline
Models & SSPs   & Ages & Metallicities & Notes  \\
\hline
{\bf No.(1)} & {\bf 6 SSPs }  & {\bf Y1 Y2; I2 I3; O3 O4 } & {\bf Z2  } &  \\
       &               &{\bf Y3 Y4; I2 I3; O3 O4  } & {\bf Z2  } & {\bf for VM }\\ 
\hline
No.(2) & 3 SSPs       & Y1; I2; O3          & Z2 & \\ 
       &              & Y3; I2; O3          & Z2  & for VM \\ 
\hline
No.(3) & 9 SSPs   & Y1 Y2 Y3; I2 I3 I4; O2 O3 O4 & Z2 &  \\
       & 8 SSPs   &    Y3 Y4; I2 I3 I4; O2 O3 O4 & Z2 & for VM \\ 
\hline
No.(4) & 12 SSPs & Y1 Y2 Y3 Y4; I1 I2 I3 I4; O1 O2 O3 O4 & Z2 & \\
       & 10 SSPs &       Y3 Y4; I1 I2 I3 I4; O1 O2 O3 O4 & Z2 &  for VM\\ 
\hline
No.(5) & 12 (6$\times$ 2) SSPs  &  Y1 Y2; I2 I3; O3 O4  & Z2 Z1  & \\
       &                     &  Y3 Y4; I2 I3; O3 O4  & Z2 Z1  &  for VM\\ 
\hline
No.(6) & 12 (6$\times$ 2) SSPs     & Y1  Y2; I2 I3; O3 O4  & Z2 Z3 & \\
       &                        &  Y3 Y4; I2 I3; O3 O4  & Z2 Z3 &  for VM\\ 
\hline
No.(7) & 18 (6$\times$ 3) SSPs     & Y1 Y2; I2 I3; O3 O4  & Z1 Z2 Z3 & \\
       &                        & Y3 Y4; I2 I3; O3 O4  & Z1 Z2 Z3 &  for VM\\
\hline
No.(8) & 6 SSPs & Y1 Y2; I2 I3; O3 O4 & Z1 & \\
       &        & Y3 Y4; I2 I3; O3 O4 & Z1 & for VM \\
\hline
No.(9) & 6 SSPs & Y1 Y2; I2 I3; O3 O4 & Z3 & \\
       &        & Y3 Y4; I2 I3; O3 O4& Z3& for VM \\
\hline
\end{tabular}
\end{table*}

\subsection{Spectral analysis and comparisons among six different EPS models}
\label{sec.compare}

Among the six EPS models, the SSPs in three of them (BC03, CB07, Vazdekis/Miles)
have similar spectral resolution, and are
comparable to the resolution of SDSS spectra;
another three (Ma05, GALEV, GRASIL) provide lower spectral
resolution SSPs, $\sim$20\AA. 
Thus, when we use SSPs in the latter three EPS models, 
we need to decrease the resolutions of the observed spectra to match the models. 
We have tried three methods to do that,
the first is the GAUSS task in IRAF, 
the second is Disgal1D developed by our French collaborators, and the third is 
re-bin program kindly provided by Dr. Jingkun Zhao at NAOC. 
We find there are no much difference among the results 
from these three methods.
Therefore, finally we adopt the GAUSS task in IRAF for 
de-resolution on our spectra.

In this section, we adopt SSP group No.(1) from 6 different EPS models 
to analysis the stellar populations. 
In that way, the selection of age, metallicity, and IMF are fixed, and the only free 
factor is EPS models. So that we can 
check the dependences of stellar population synthesis on EPS models. 
Take star-forming and E+A galaxies as examples, 
we analyze their stellar populations by using group No.(1) from 
these six EPS models, and then compare the results. 

 \begin{figure*}
 \centering
 \includegraphics[width=6cm]{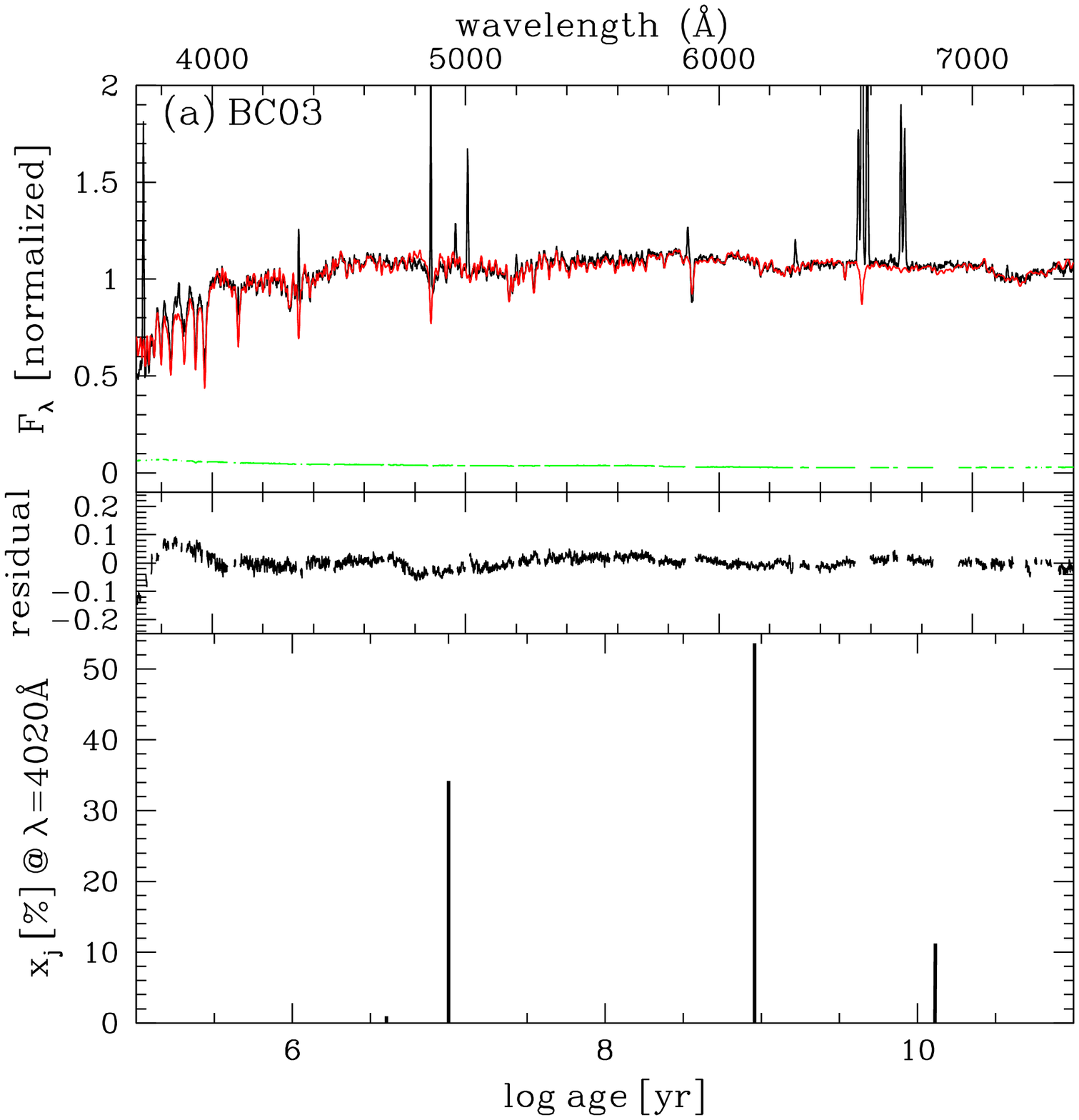}
 \includegraphics[width=6cm]{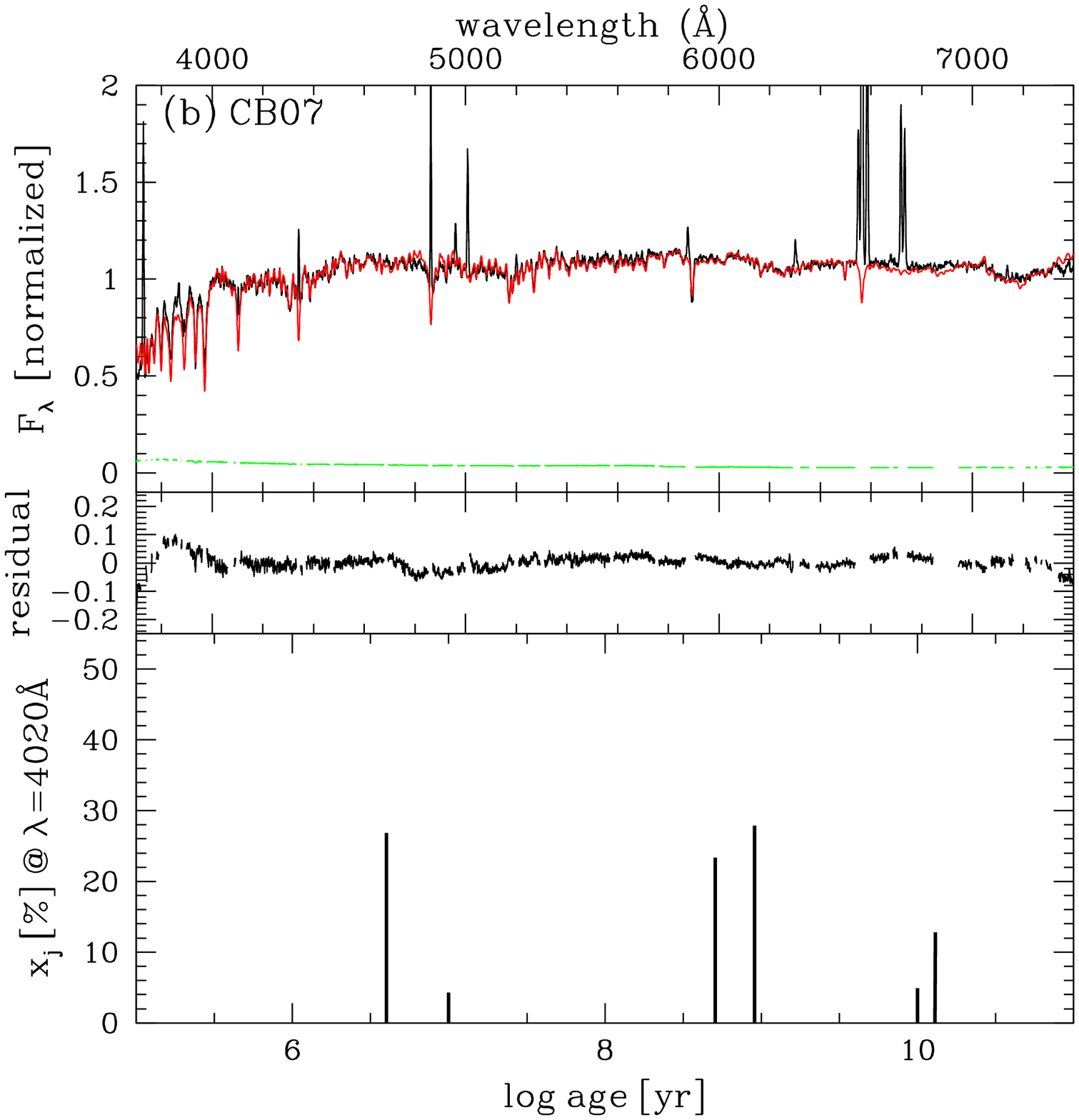}
 \includegraphics[width=6cm]{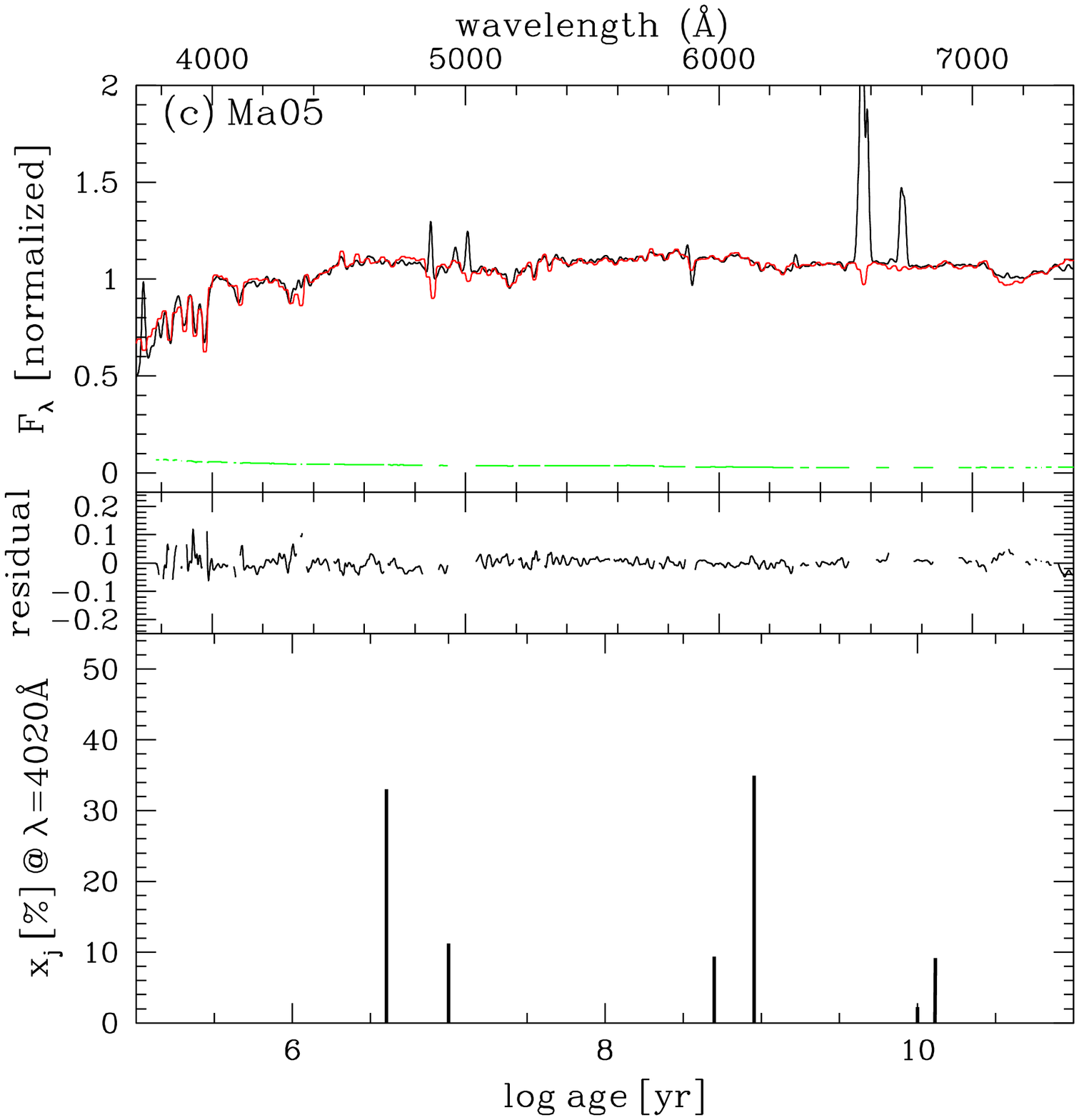}\\
 \includegraphics[width=6cm]{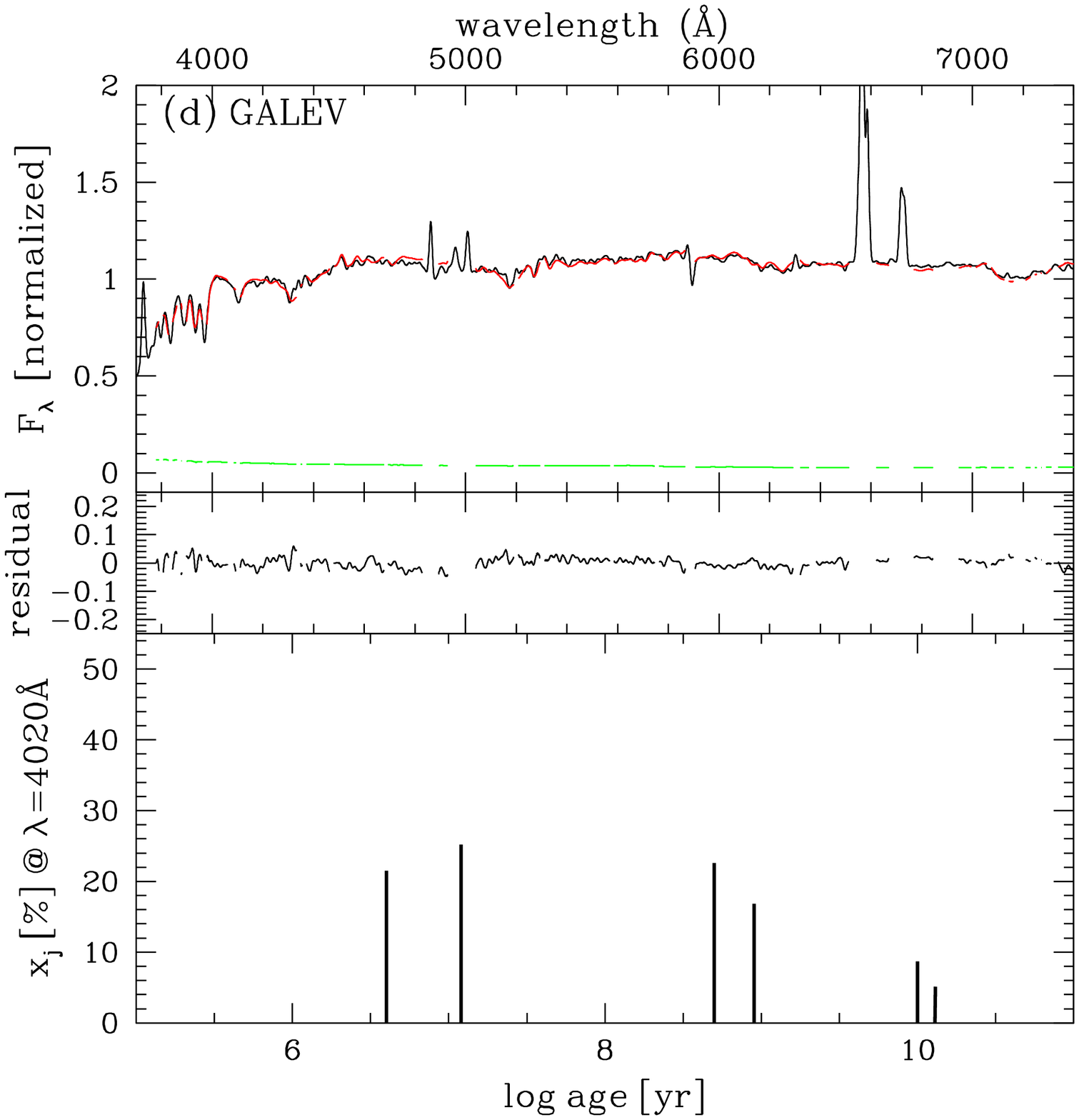}
 \includegraphics[width=6cm]{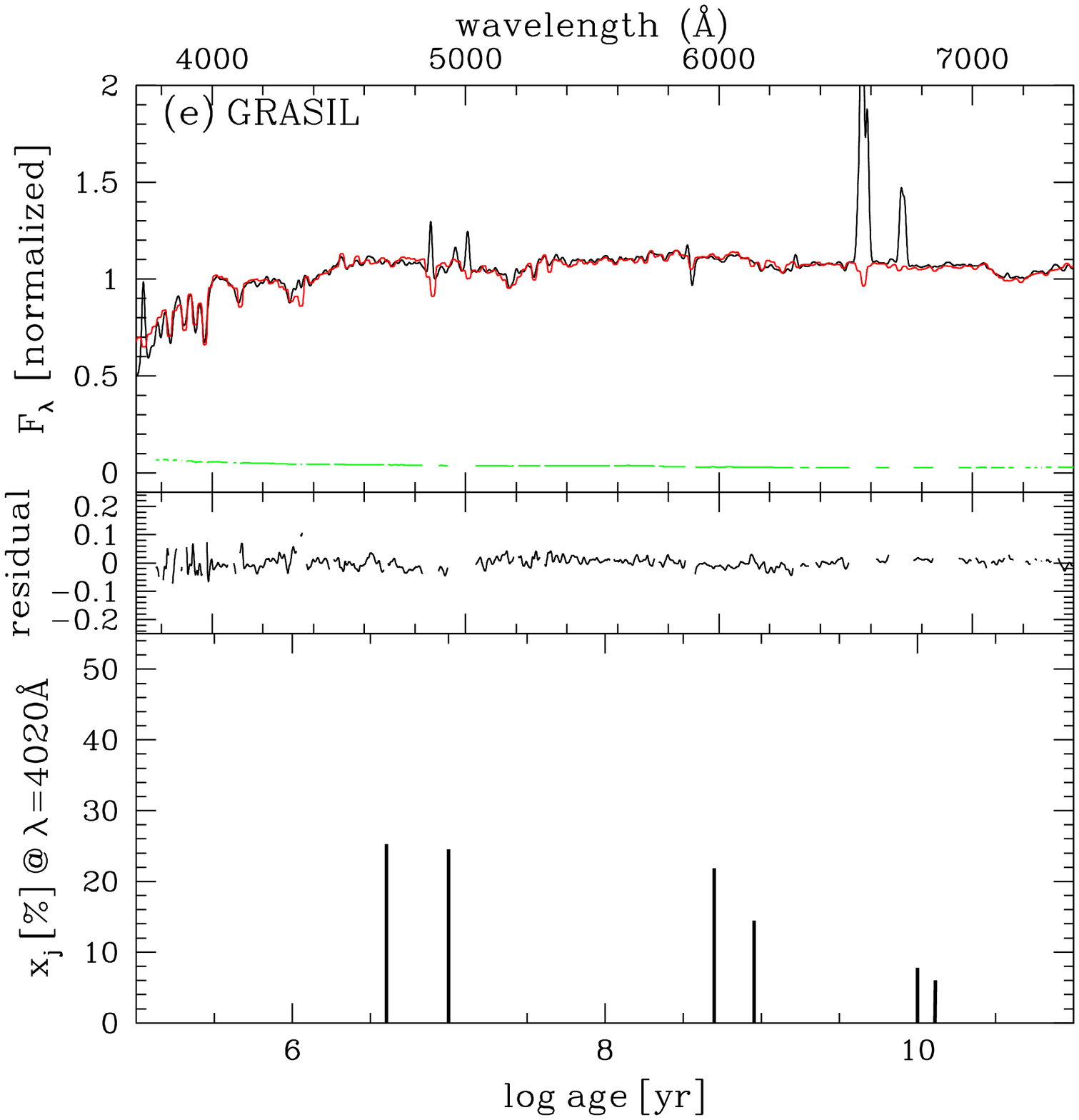}
 \includegraphics[width=6cm]{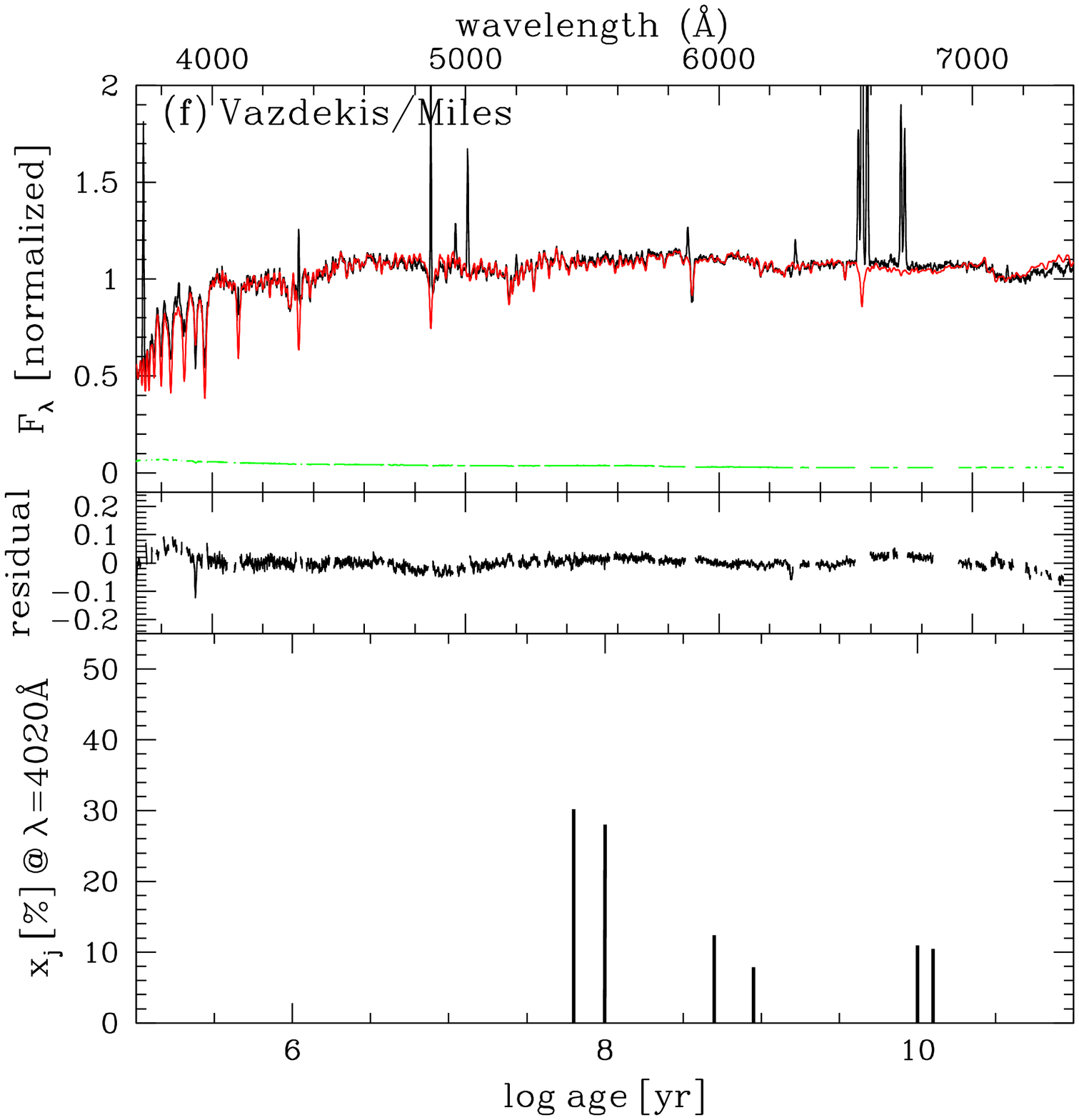}
\caption{Spectral fitting results of star-forming galaxies by using six SSPs (group No.(1)) 
from different 
models: (a) BC03, (b) CB07, (c) Ma05, (d) GALEV, (e) GRASIL and (f) Vazdekis/Miles. 
In each EPS model, 
top panels: comparison of synthesis spectrum (red line) with the observed spectrum 
(black line), and green line shows the error spectrum; middle panels: the residual spectrum; 
bottom panels: distributions of light fractions. 
}
\label{sf.allmodels.6ssp}
\end{figure*}

\begin{figure*}
\centering
\includegraphics[width=6cm]{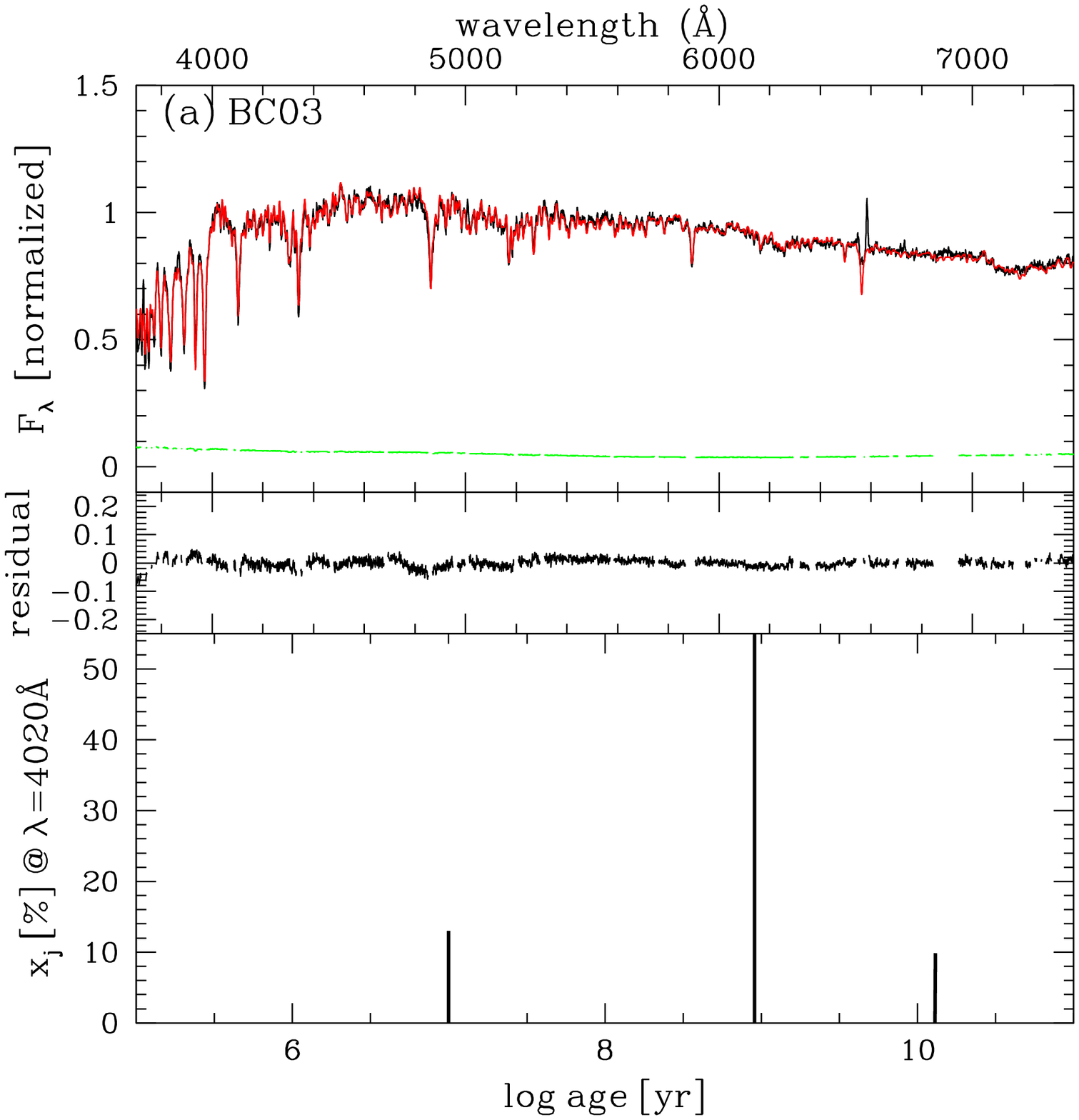}
\includegraphics[width=6cm]{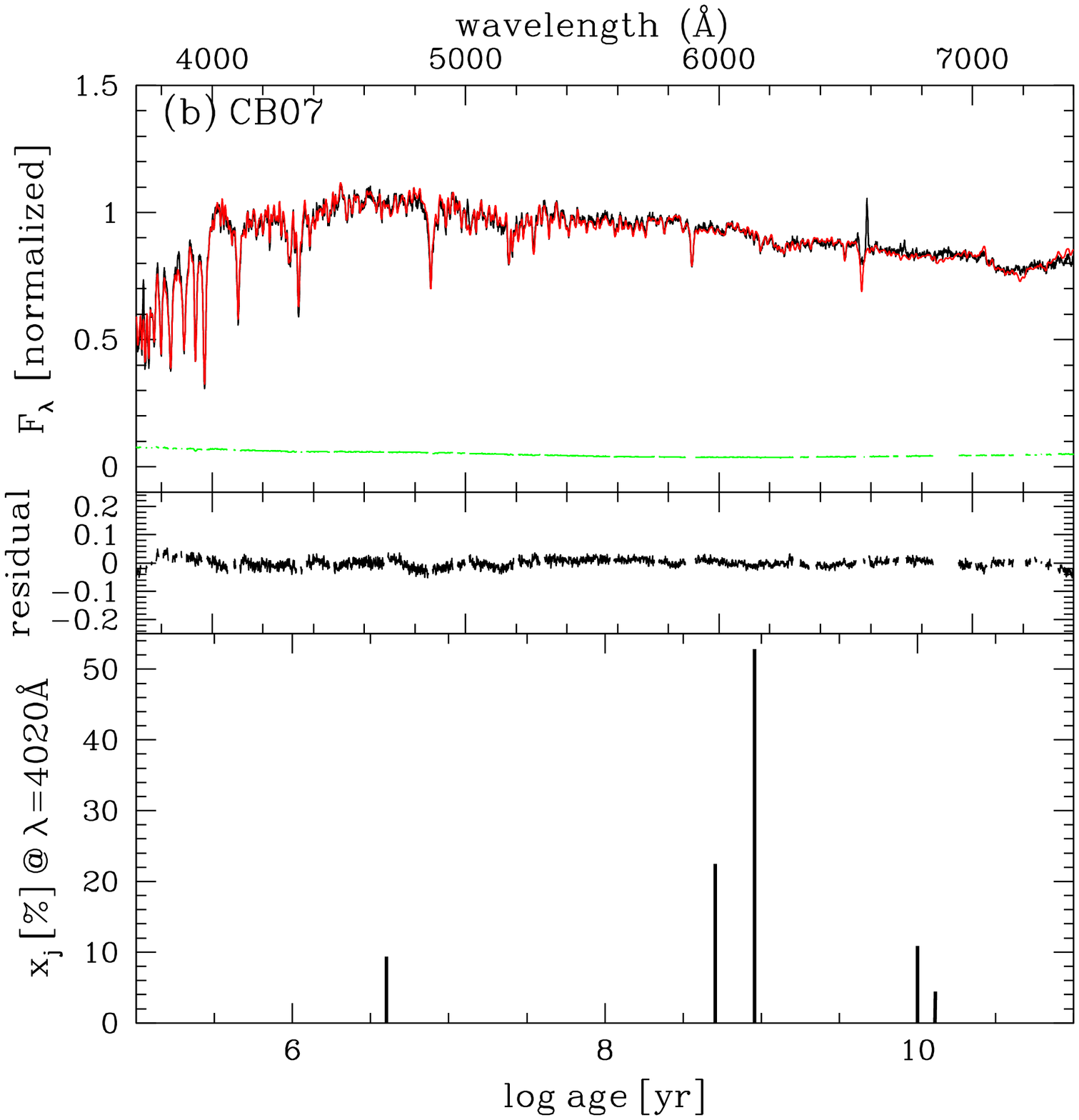}
\includegraphics[width=6cm]{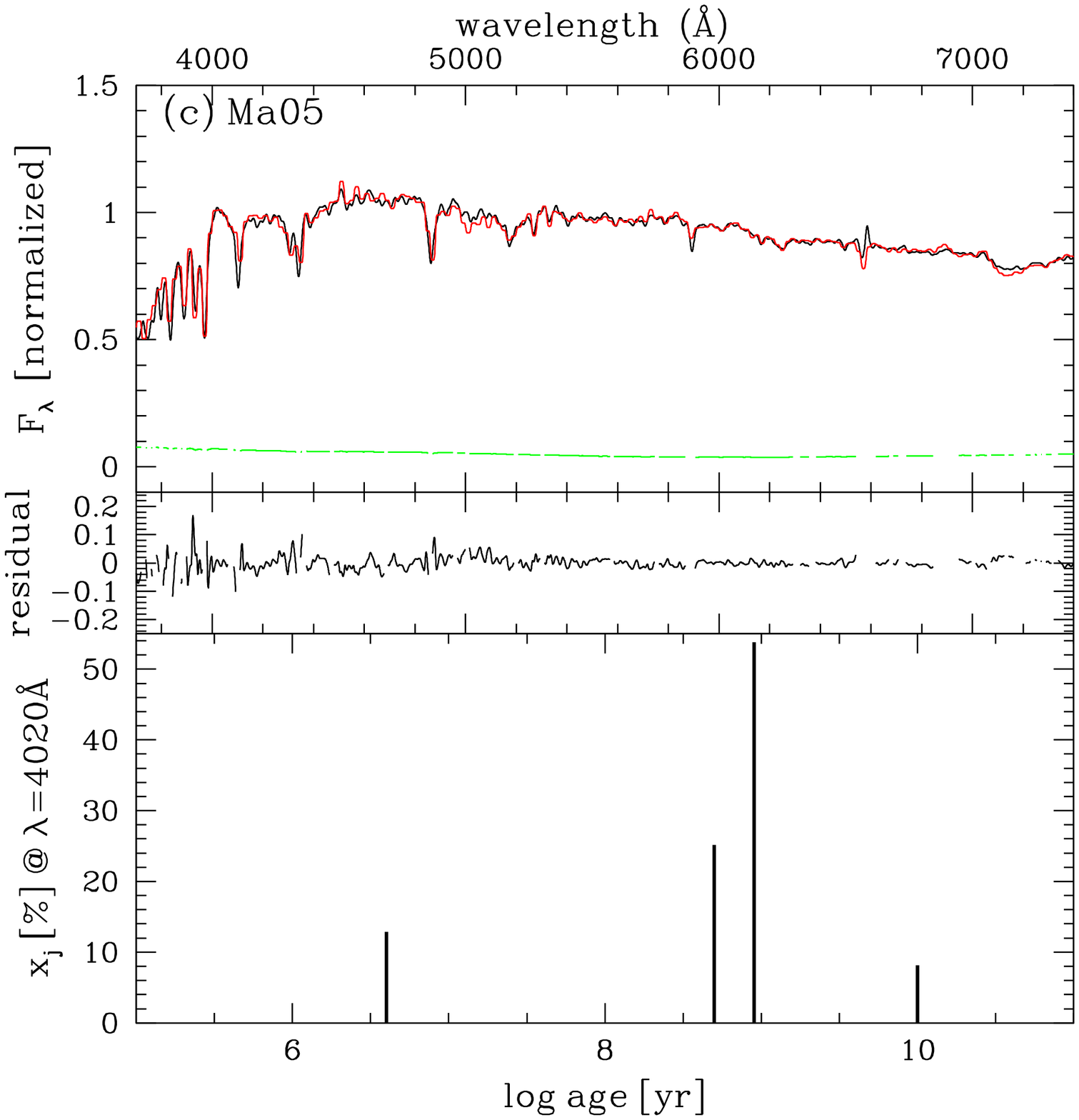} \\
\includegraphics[width=6cm]{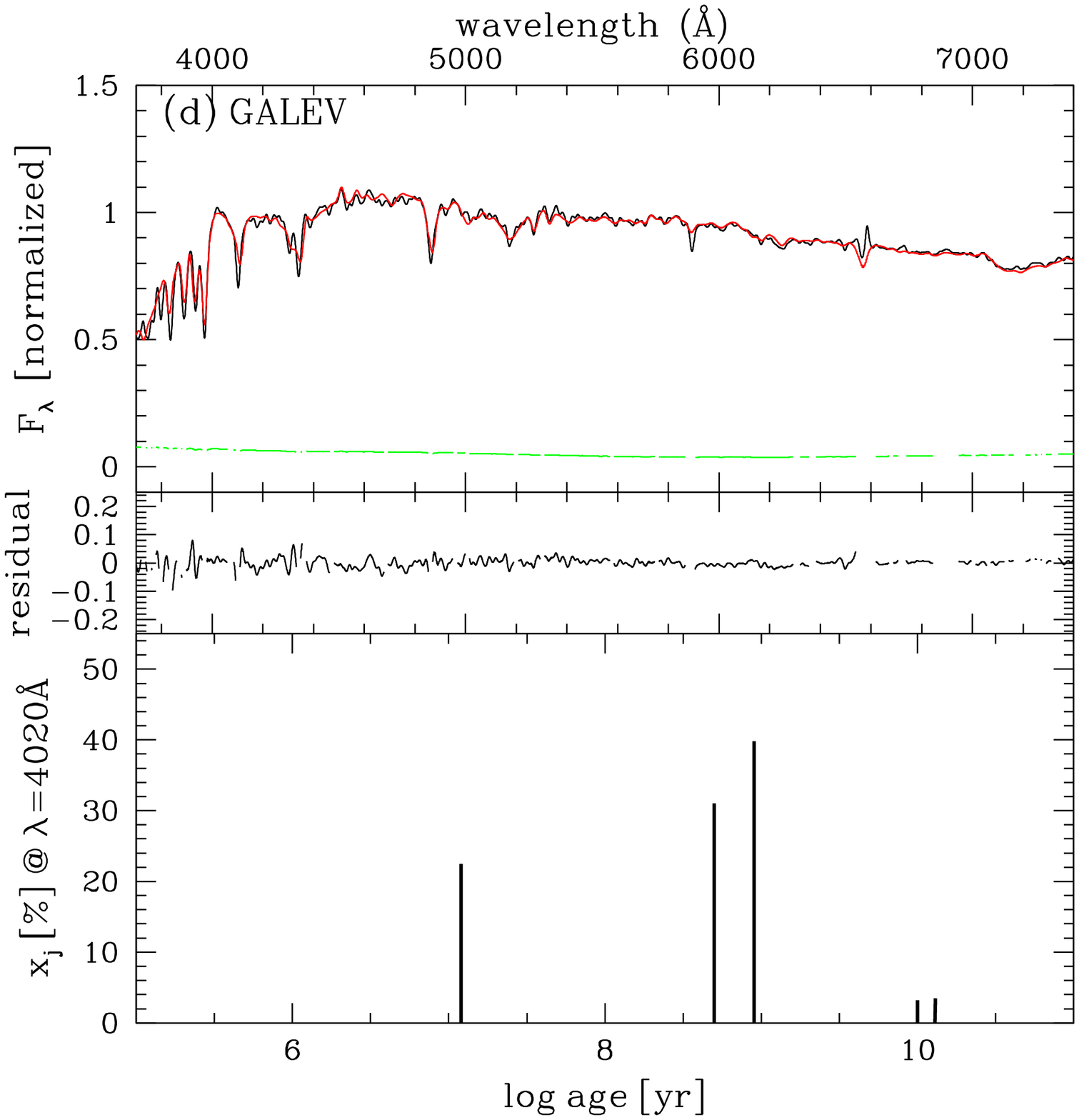}
\includegraphics[width=6cm]{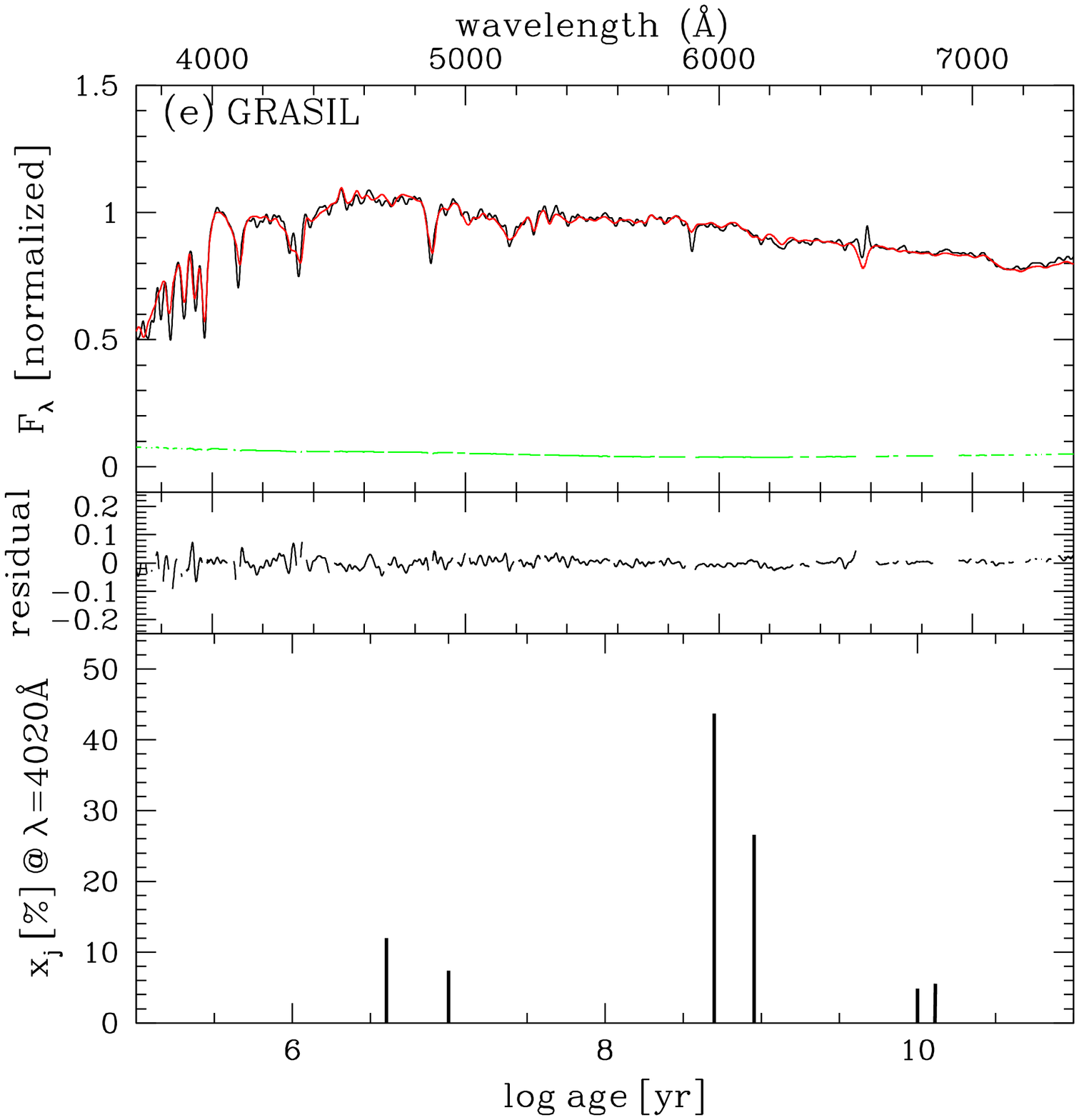}
\includegraphics[width=6cm]{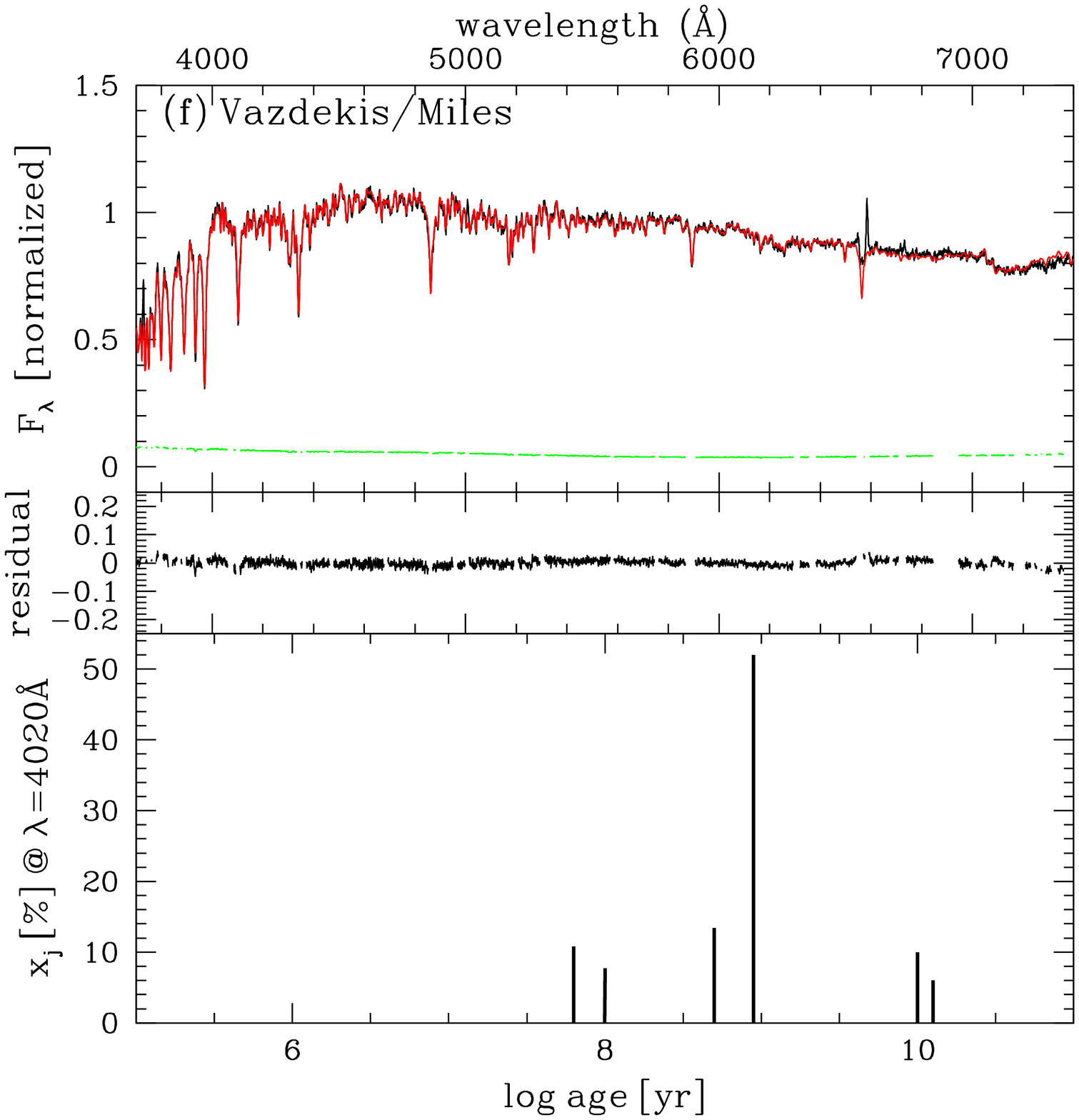}
\caption{Same as Fig.~\ref{sf.allmodels.6ssp}, but for E+A galaxies.
}
\label{kpa.allmodels.6ssp}
\end{figure*}

\begin{table*}
\caption{Stellar populations in star-forming and E+A galaxies by using different 
EPS models with SSP group No.(1) (Tables~\ref{agez.tab},~\ref{groups.tab}). 
Y represents young populations, I represents intermediate populations, 
O represents old populations. Please see details of 
age and Z selections for each SSP group in text. 
}
\centering
\label{sk.6ssp.allmodels}
\begin{tabular}{l|l|c|c|c|c|c|c|c}
\hline
Types & SSP groups & Age bin & BC03& CB07& Ma05& GALEV & GRASIL & Vazdekis/Miles\\
\hline
star- &  No.(1) &  Y&  35   &  31 & 44 & 47 &  50 &  58 \\
forming &  6 SSPs &  I&  54 &  51 &  44 &  39 &  36 &  20 \\
galaxies &  Z=Z2  &  O&  11&  18 &  12 &  14 &  14 &  22 \\
\hline \hline
E+A &   No.(1) &  Y         & 13 &9 & 13 & 22 & 20 & 19 \\
galaxies &   6 SSPs   &  I   & 77& 75 & 79 & 71 & 70& 65 \\
 &   Z=Z2       &  O         & 10 & 15 & 8 & 7 & 10    & 16 \\
\hline
\end{tabular}
\end{table*}

\subsubsection{Star-forming galaxies}
\label{sec.sf}
In Fig~\ref{sf.allmodels.6ssp}
we present the spectral fitting results and the light fraction distributions 
for the combined spectra of star-forming galaxies 
by using six SSPs (group No.(1) in Table~\ref{groups.tab}) 
from six different EPS models. 
Every three panels as a group correspond to the results from one EPS model. 
The top panel shows the observed spectra (black line), 
the synthesis spectra (red line), and the error spectra (green line); 
the middle one shows the residual spectra; 
and the bottom one shows the light fraction distribution. 
From the top two panels in each model, 
we can see that the synthesis spectra match well with the observed spectra in most cases.
However, one point we would like to notice is, 
there is a trough around $H{\beta}$ when we use BC03 and CB07, 
and it disappears when we use Ma05, GALEV, GRASIL and Vazdekis/Miles. 
Asari et al. (2007) suggested that this could relate to calibrations in the STELIB library 
in this spectral range, and we also confirmed it in our earlier work (Chen et al. 2009) by 
using star clusters. 
We speculate that the bad fit may also be due to the way to handle the shape of 
the continuum in Starlight. To check this point, we further tried another program, ULySS 
(Koleva et al. 2009b), 
to perform similar fits, and we got much better fits around $H\beta$. However, 
this problem does not affect much on our stellar population analysis.
Besides that, we find the Ma05, GALEV and GRASIL models seem to be
superior in  reproducing the important Balmer lines and continuum. 
This could be due to the decreasing resolution, which reduces the details in
spectra. 

The bottom panels in Fig~\ref{sf.allmodels.6ssp} show that 
the resulted stellar populations from different EPS models
are not exactly same for the star-forming galaxies, 
although the dominant populations are all young plus intermediate populations.  

The top line of Table~\ref{sk.6ssp.allmodels} presents the contributed light fractions from 
the young (Y), intermediate (I) and old (O) populations.
The related numbers reveal that star-forming galaxies are composite of young, 
intermediate and old populations. 
The differences of the contributed light fractions
among these EPS models are in a range of
 0.01\% $\sim$ 18.75\%.
Vazdekis/Miles is different from others, and it results
in 58\% (27\% more than that of CB07) young and 20\% intermediate populations. 
The reason of this could be  
Vazdekis/Miles has no enough young populations as others,
and its youngest age is Y3=0.063 Gyr, older than others,
which may result in higher fraction of young population. 
We believe that the error-bars of these resulted light fractions are small. 
As mentioned by Cid Fernandes et al. (2005) that 
they presented the error-bars centered on the mean values obtained by fitting 
20 realizations of each of 65 test galaxies. And their three condensed populations 
are well recovered by Starlight, with uncertainties smaller than 
0.05 (young: $t<10^{8}$), 0.1 (intermediate: $10^{8}<t<10^{9}$), 
and 0.1 (old: $t>10^{9}$) for S/N $\geqslant 10$ respectively. 
In our fittings, the code provides the values of last-chain-values for 7 chains, and 
we find that most of the discrepancies in these adopted values are less than 1\%. 
So we will not consider error-bars in our following studies.

Moreover, it also shows that BC03 and CB07 result in 
$\sim$30\% young stellar population and 
$\sim$50\% intermediate age one.
Ma05, GALEV and GRASIL
result in comparable young and intermediate populations 
(both $\sim$44\% in Ma05) or slightly higher young populations
than the intermediate ones ($\sim$47\% vs. 39\% in GALEV; 
$\sim$50\% vs. 36\% in GRASIL). 
In other words, there
is an inversion of the dominant population between BC03, CB07 on one side, and
GALEV, GRASIL and Vazdekis/Miles on the other side. 
In order to explain this phenomenon, 
we directly compare the 6 SSPs from these models, which we used in the fittings. 
And we find the young SSPs of BC03 and CB07 hold relatively 
higher fluxes at blue band than others when we normalize them
at around 6000\AA. Take Y2 as an example, GALEV and GRASIL as one group, the fluxes at 
blue band of which are lower than that of BC03 and CB07, and Ma05 is between 
these two groups. This explains the inversion of 
young and intermediate populations between the two groups, 
and further Ma05 produces comparable young and intermediate ones. 
As for Vazdekis/Miles, the two SSPs in young age-bin are older than others, 
thus the resulted young population is much higher than the intermediate one. 
The reason to cause such difference in SSPs among these 
EPS models could be their different stellar libraries. 
STELIB is used by BC03 and CB07, 
but BaSeL2.0 or Kurucz are used in GALEV, Ma05 and GRASIL. 
The former is empirical from real observed stars' spectra with higher resolutions 
and the latter are theoretical ones with lower resolutions. 
The limit of STELIB library
has been fully discussed by Koleva et al. (2008) and 
Gonzalez Delgado \& Cid Fernandes (2009). 
Another possible reason causing such difference between the two groups 
could be the spectral resolution. Since BC03 and CB07 have higher resolutions (3\AA), while 
Ma05, GALEV and GRASIL have lower resolutions (20\AA). 
To check this factor, we degraded the 6 SSPs in BC03 from 3\AA to 20\AA, and 
then performed similar fits. 
We find that the discrepancies between the two groups decrease then.

\subsubsection{E+A galaxies}
\label{sec.kpa.ssp}
In Fig.~\ref{kpa.allmodels.6ssp} we perform the spectral fitting results and the light fraction distributions 
of E+A galaxies by using six SSPs 
(group No.(1) in Table~\ref{groups.tab}) from six different EPS models.

We find that the matching between 
the synthesis spectra and the observed spectra is good. 
Meanwhile, the troughs around $H_{\beta}$ 
become shallower associated to BC03 and CB07 than that of star-forming galaxies, 
and disappear in panels corresponding to Ma05, GALEV, GRASIL and Vazdekis/Miles, which 
further confirm the conclusion we draw in studying star-forming galaxies. 
Similarly, the bottom panels show that different EPS models do not give 
exactly same populations for E+A galaxies,  
even though the peaks of distributions are more concentrated. 

In the bottom line of Table~\ref{sk.6ssp.allmodels},
we also list the resulted light fractions 
of E+A galaxies.  
It shows that the contributions from different models vary within a range of 
0.14\% $\sim$ 13.53\%. Moreover, the percentages of intermediate-age populations 
are all greater than 70\% except for Vazdekis/Miles (65.45\%). 
Thus we suggest that E+A galaxies are 
dominated by intermediate-age populations, and are composed of 
young, intermediate-age and old populations. 

In all, to sum up the points about stellar population analysis among different EPS models which 
we have just indicated, 
among different EPS models the numerical values of light fractions change obviously, 
although the dominant populations are consistent.
Additionally, it seems that EPS models still have difficulties in disentangling stellar 
populations, i. e. for a given galaxy, different models would give different order of 
importance of old, intermediate and young stellar population (especially in the case of 
star-forming galaxies). We suggest that the use of near-IR photometry maybe a way to improve 
this. 
However, as mentioned by Eminian et al. (2008), Maraston (2005) and Conroy et al. 
(2009a), TP-AGB stars dominate the near-IR light of galaxies. So we should be careful that the 
uncertainties associated with the TP-AGB phase will probably hamper significant improvements. 

 \begin{figure*}
 \centering
 \includegraphics[width=6cm]{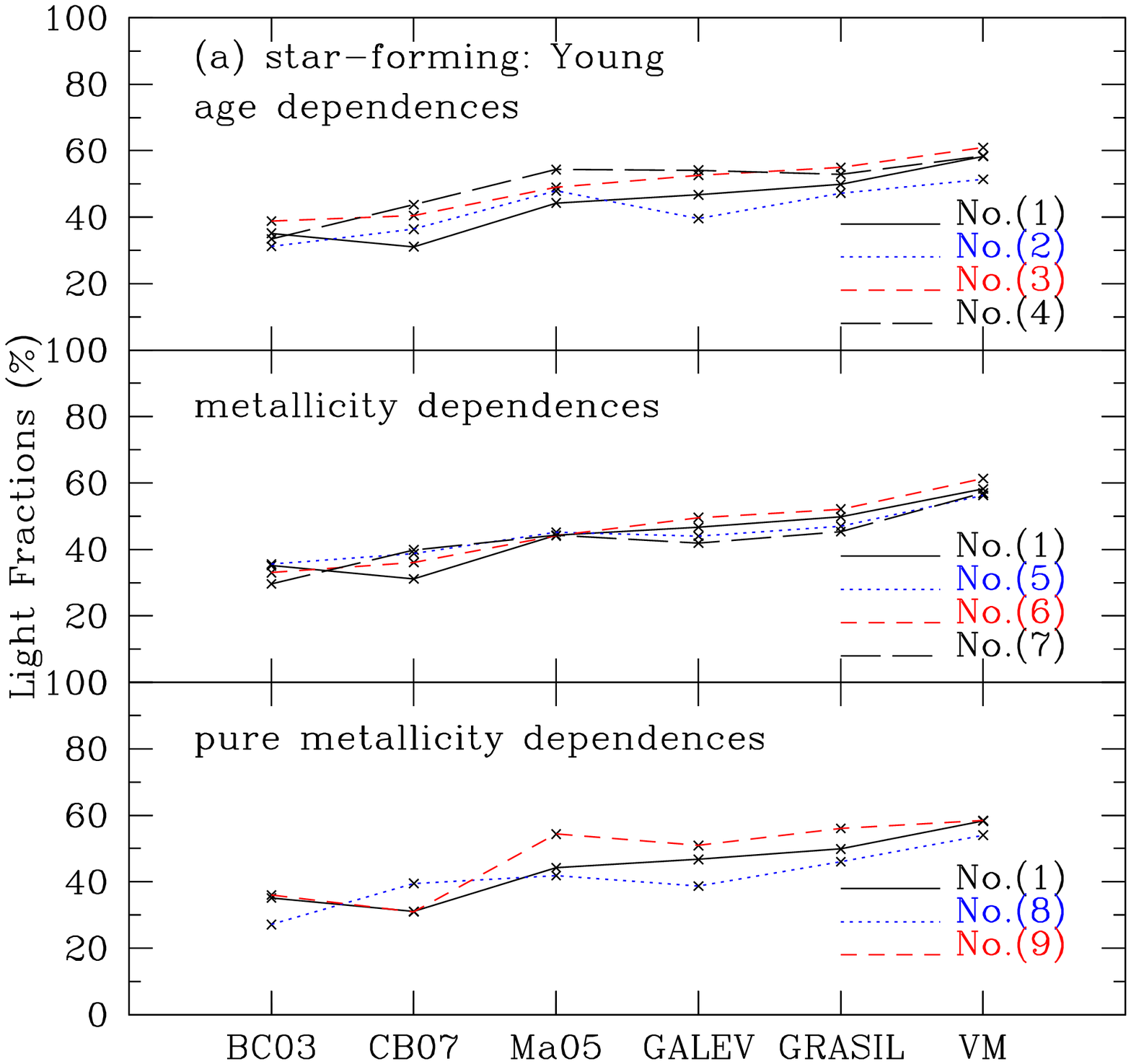}
 \includegraphics[width=6cm]{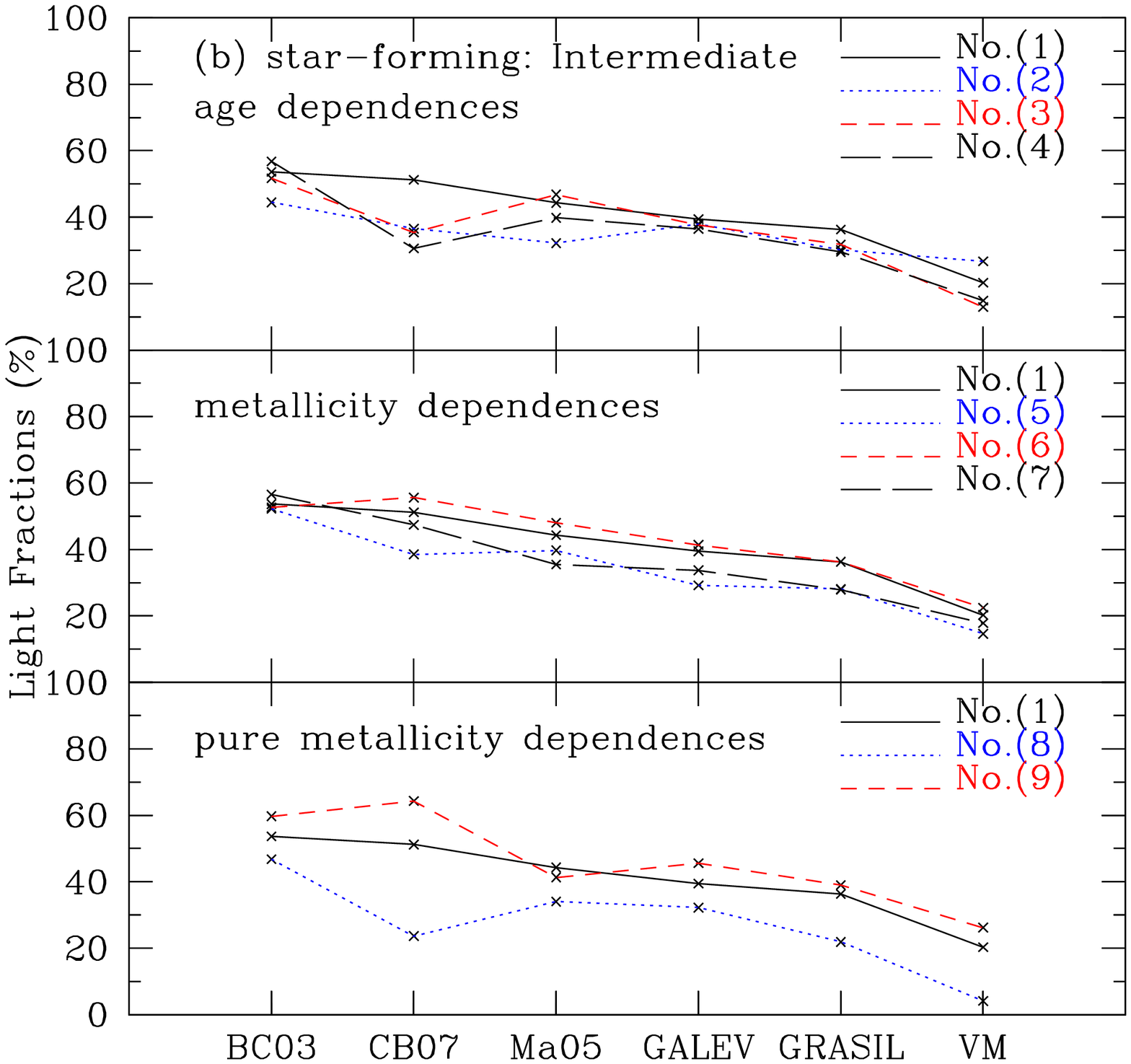}
 \includegraphics[width=6cm]{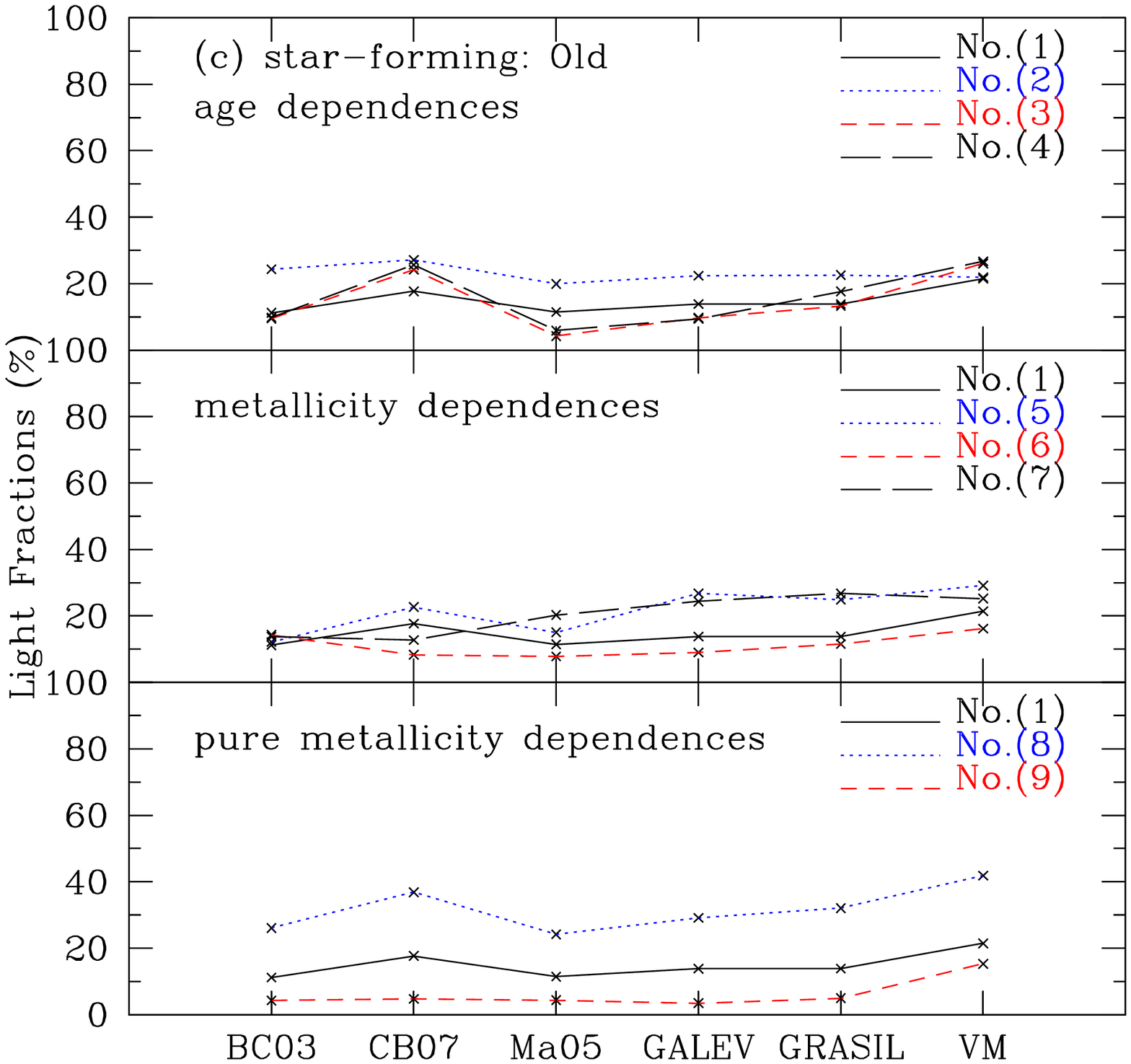}
\caption{The light fractions' contributions to star-forming galaxies in bins of 
young population (left three panels), intermediate-age population (middle three panels), and 
old population (right three panels). The x-axis represents six EPS models, and the y-axis 
represents the light fractions in percentage. Different lines connect points from 
different SSP groups. 
Top panels: solid line stands for group No.(1), dotted line stands for group No.(2), 
short-dashed line stands for group No.(3), and long-dashed line stands for group No.(4). 
Middle-panels: solid line stands for group No.(1), dotted line stands for group No.(5), 
short-dashed line stands for group No.(6), and long-dashed line stands for group No.(7). 
Bottom-panels: solid line stands for group No.(1), dotted line stands for group No.(8), 
short-dashed line stands for group No.(9).
}

\label{sf.yio.eps}
\end{figure*}

 \begin{figure*}
 \centering
 \includegraphics[width=6cm]{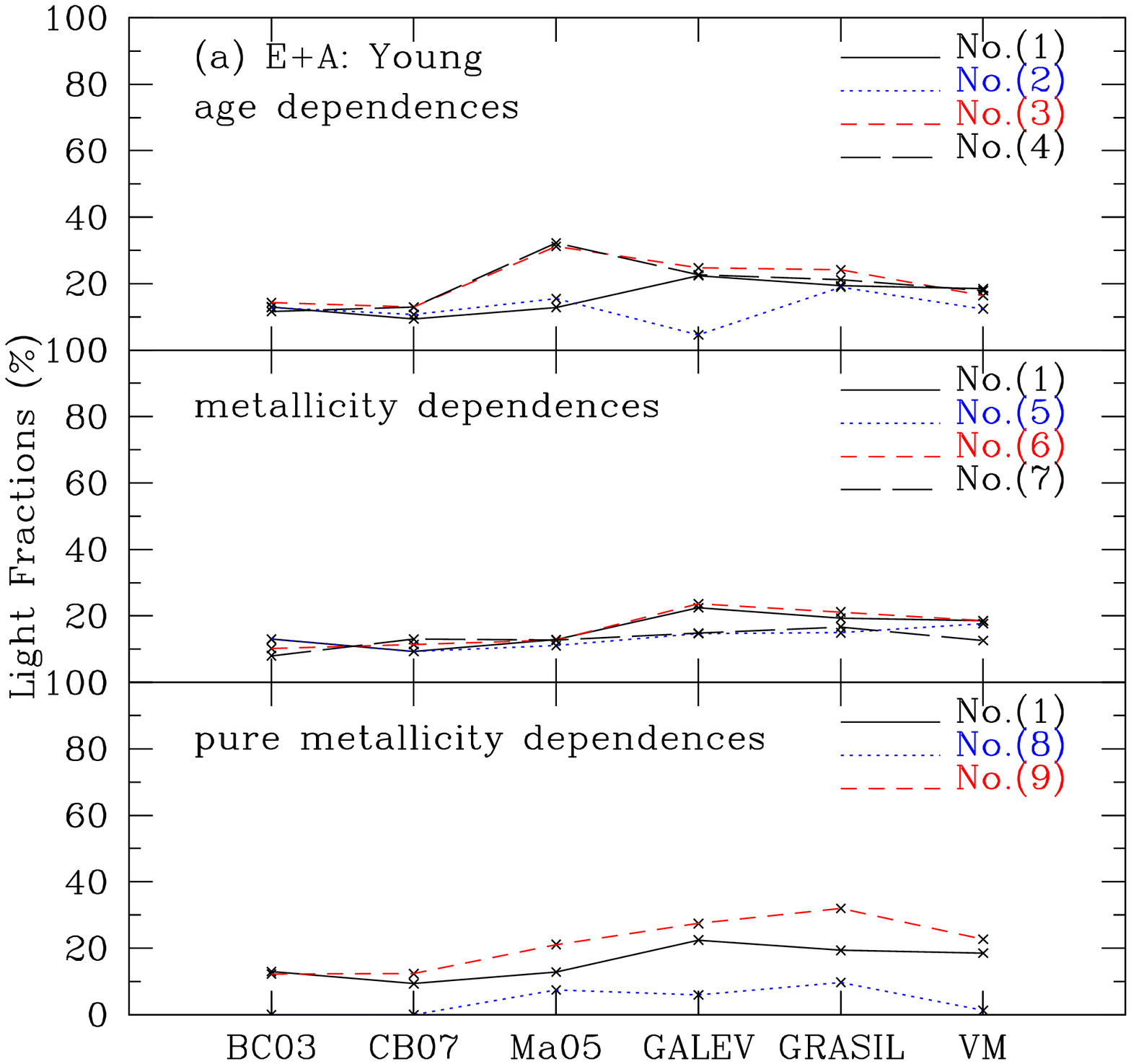}
 \includegraphics[width=6cm]{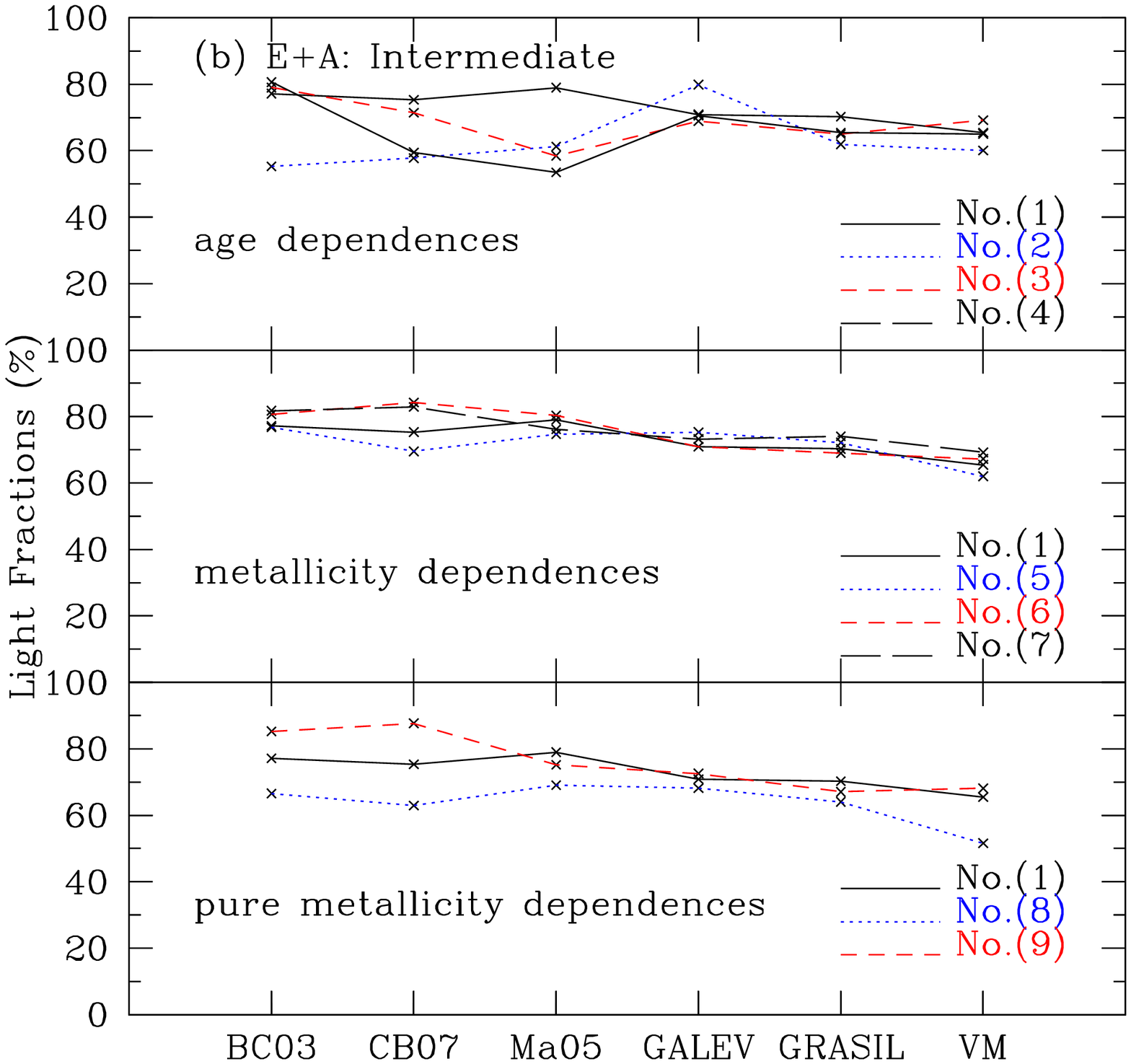}
 \includegraphics[width=6cm]{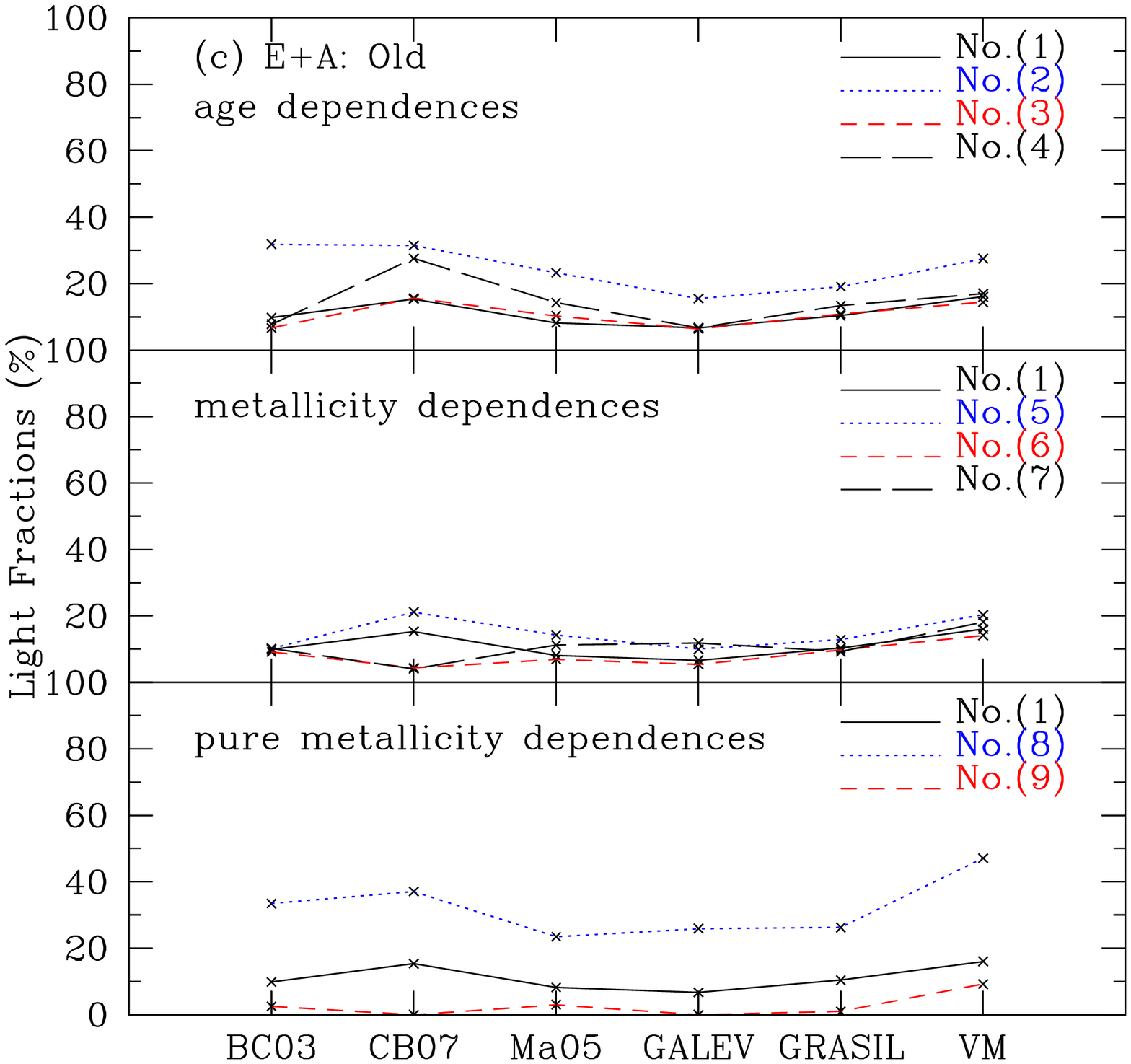}
\caption{Same as Fig.~\ref{sf.yio.eps}, but for E+A galaxies. 
}
\label{kpa.yio.eps}
\end{figure*}

\subsection{Dependences on ages and metallicities}
\label{sec.agez}
In this section we will investigate the dependence of 
stellar population synthesis on the selection 
of age and metallicity. 
As mentioned in Sect.~\ref{sec.ssp}, Table~\ref{agez.tab} and 
Table~\ref{groups.tab}, we adopt nine SSP groups in total, 
No.(1)-(9). 

\subsubsection{Age dependences - SSP groups No.(2)-(4) in Table~\ref{groups.tab} }

As mentioned above, we fix the metallicity as solar one, then we only change the selections 
of age to construct SSP groups No.(2)-(4). 
Then we use SSP groups No.(2)-(4) from different EPS models to analyze  
 the stellar populations of star-forming and E+A galaxies.
 The resulted stellar populations are compared with each other
 and also with group No.(1), 
 so that we can check the age effects on stellar population analysis.
 
The top panels in Fig.~\ref{sf.yio.eps} show the light fractions 
of young (top-left panel), intermediate (top-middle panel), and old (top-right panel) 
populations for star-forming galaxies by using the 6 EPS models.
In each panel, the horizontal axis represents the 6 EPS models, 
and the vertical axis represents the contributed light fractions. 
The solid lines connect the resulted light fractions corresponding to group No.(1). 
The dotted, dashed and long dashed lines
refer to the connections of the resulted light fractions
from the SSP groups No.(2), (3) and (4), respectively. 

From these comparisons, generally we find that 
adding more SSPs within each age bin (based on the ``standard case") 
results in more young populations(see the top-left panel), 
but removing SSPs does not result in consistent trend for them. 
In each EPS model, 
most of the changes in young populations (top-left panel) 
are $\sim$1-10\% among different SSP groups. 
The biggest changes occur between the
groups No.(1) and No.(4) in CB07 ($\sim$16\%) as well as between groups 
No.(2) and No.(4) in GALEV ($\sim$15\%).
As to the general trend of 
intermediate-age populations (the top-middle panel), 
adding/removing SSPs results in less contributions. 
The most obvious changes  
occur between the groups No.(1) and No.(4) in CB07 ($\sim$20\%)
and between No.(2) and No.(3) in Ma05 ($\sim$14\%). 
The changes in old populations (the top-right panel) are 
related to the young and intermediate 
ones, since the sum of these three populations is equal to 100\%.

Similarly, we plot the corresponding results for E+A galaxies in 
top panels of Fig.~\ref{kpa.yio.eps} to check the age dependences.
The symbols and lines have the same meanings as in Fig.~\ref{sf.yio.eps}.
It confirms the general trend in star-forming galaxies, i.e. 
adding more SSPs increases the young populations and
decreases the intermediate populations. 
And 3 SSPs result in complex changes for the different models.
 In details, the top-left panel shows that within each EPS model, most of the changes 
in young populations are around 0-5\% among different SSP groups. 
The biggest ones occur between groups No.(1) and No.(4) in Ma05 
as well as between groups No.(2) and No.(3) in GALEV ($\sim$ 20 \%); 
the changes in intermediate-age
 and old populations are obvious in CB07 and Ma05, up to $\sim$ 18\%
 and $\sim$ 25\%. 
 The reason could be that
 CB07 and Ma05 specially include the TP-AGB contributions to build their SSPs, 
 and the E+A galaxies are right dominated
 by intermediate-age populations, which are quite related to the TP-AGB stars.
 This could also explain the most obvious discrepancies in the
 intermediate-age populations of star-forming galaxies with different
 models (see top-middle panel of Fig.~\ref{sf.yio.eps}).  
Moreover, the obvious discrepancy of group No.(2) with 3 SSPs from others may mean
that 3 SSPs are not enough for such stellar population analysis on galaxies,

In summary, from the top panels in Figs.~\ref{sf.yio.eps}, \ref{kpa.yio.eps}, 
we can comment that the SSPs with different sets of ages present 
obviously different results (such as CB07, Ma05), 
although the discrepancies from both the models and ages
will not change the dominant population of the galaxies.

\subsubsection{Metallicity dependences - SSP groups No.(5)-(9) in Table~\ref{groups.tab} }

We fix the selections of age as the same as in the group No.(1) 
(Table~\ref{groups.tab}, solar metallicity), 
then we add sub-solar metallicity or/and super-solar metallicity to construct 
groups No.(5)-(7) (Table~\ref{groups.tab}).
By comparing the results from these SSP groups with those from group No.(1), 
we can check the metallicity effects on 
spectral synthesis results. 
 
The middle panels in Fig.~\ref{sf.yio.eps} show the 
light fractions' 
contributions from
young (middle-left panel), intermediate (middle-middle panel), and old (middle-right panel) 
populations to star-forming galaxies by using the 6 EPS models.
The solid lines connect the resulted light fractions corresponding to group No.(1) 
from different EPS models. 
The dotted, short dashed and long dashed lines
refer to the connections of the resulted light fractions
from the SSP groups No.(5), (6) and (7), respectively.

These results show that within each EPS model 
all of the changes in young populations are less than 5\%. 
The changes in intermediate-age populations (middle-middle panel) are 
larger and are in a range of 0-17\%, such as
$\sim$ 17\% between groups No.(5) and No.(6) in CB07,
$\sim$ 13\% between groups No.(5) and No.(6) in Ma05. 
For the old population (middle-right panel), 
the changes are also obvious among these SSP groups (0-17\%),
e.g., $\sim$ 14\% between groups No.(5) and No.(6) in CB07,
$\sim$ 17\% between groups No.(5) and No.(6) in GALEV.
Similarly, the more obvious changes in CB07 and Ma05 may be related to the 
involved TP-AGB stars in their SSPs.

Similarly, we present the corresponding results for E+A galaxies in the middle panels in 
Fig.~\ref{kpa.yio.eps}. It shows that within each EPS model, all changes in 
young populations are not obvious (less than 6\%); 
generally the changes in intermediate-age populations and old populations are also
not obvious except for CB07. 
The differences is about 15\% between groups No.(5) and (6) in intermediate
population, and about 15\% between 
groups No.(5) and (7) (as well (6)) in old population (within CB07).
These could also be due to the TP-AGB effect. 

In a word, from the middle panels in Figs.~\ref{sf.yio.eps}, \ref{kpa.yio.eps}, 
we suggest that SSPs with different selections of metallicities 
give different results. 

By comparing the top panels with the middle panels in Figs.~\ref{sf.yio.eps},
\ref{kpa.yio.eps}, 
We can see that the effects of metallicity is less than ages. 
However, this phenomenon may be due to either the way we show our results 
or the age-metallicity degeneracy. 
Note that up to now, we focused on the contributions of light fractions. 
While other relevant aspects, such as M/L, maybe sensitive to metallicity. 
Maraston (2005) has shown some parameters' dependences on metallicity, 
for example: the IR indices $C_{2}$ in their figure~16, the $M^{\ast}/L$ in their 
figure~23, the $D_{4000}$ in their figure~25, and the indices $CaT^{\ast}$, CaT and PaT 
in their figure~26. 
Longhetti \& Saracco (2009) also performed the $M/L$ as a function of metallicity 
in their figures~$3\sim 5$. 

Another notice is that there could be age-metallicity degeneracy following our 
adding or removing SSPs in the top and middle panels in 
Figs.~\ref{sf.yio.eps},~\ref{kpa.yio.eps}. 
For this reason we constructed the last two SSP groups, in which we only changed the 
metallicities and fixed all the ages. 
We present the results of these two groups in the bottom three panels in 
Figs.~\ref{sf.yio.eps},~\ref{kpa.yio.eps}. 
From the bottom three panels we see a much larger scatter (up to 40\%) 
among different SSP groups than 
that of middle panels. 
Such that we have to point out the age-metallicity degeneracy exists in our 
Figs.~\ref{sf.yio.eps},~\ref{kpa.yio.eps}. 
In our current paper, we will not do more work on this classic degeneracy between 
age and metallicity, and some related studies can be found in Cid Fernandes et al. (2005), 
Eminian et al. (2008), Carter et al. (2009), Cid Fernandes \& Gonz\'{a}lez Delgado (2009).

\section{Dependences on evolution track, stellar spectral library, 
and the age sequence of galaxies}
\label{sec.other}

In this section, we further check the effects of stellar 
evolution tracks 
on the stellar population analysis of galaxies.
We take BC03 and GALEV models as examples to do such check.
We will use these two models to analyze the stellar populations of
all the six representative samples of galaxies, i.e.
the star-forming, composite galaxies, Seyfert 2s, LINERs, E+A and early-type galaxies. 
Hence we can compare the different stellar populations in each galaxy type. 

\subsection{Dependences on stellar evolution track and stellar spectral library}

We use SSP group No.(1) (in Table~\ref{groups.tab}) with Salpeter IMF, 
and with two different sets of stellar evolution tracks (Pa94 and Pa00 in BC03;
Pa99 and Geneva in GALEV).
The contributed light fractions of young, intermediate and old populations 
are listed in Table~\ref{bc03.alltype.atxt}.
The first two columns corresponding to BC03 shows 
that within each galaxy type, among different choices of 
stellar evolution tracks, 
the corresponding changes in young populations are
0-4\%, in intermediate populations are  
0-9\%, and in old populations are 0-10\%.
While most of the results of GALEV (last two columns) are similar to BC03, except for 
the changes in intermediate and in old populations of early type galaxies ($~20\%$). 
These results imply that changing stellar evolution track will not change much 
the resulted stellar populations of all the six types of galaxies. 

Additionally, we mentioned in Sect.4.3.1 that 
different selections of libraries may be another factor effects our results. 
Therefore, we compared the two different libraries provided by BC03 
(Table~\ref{bc03.alltype.atxt}). 
One library is STELIB and another is BaSeL3.1, and we fix the IMF as Salpeter and 
the track as Pa94. 
We find that the results of BaSeL3.1 is different from that of STELIB, but 
the results of BaSeL3.1 is similar with that of GRASIL and GALEV. 
This supports our previous discussions in Sect.4.3.1. 
Besides our conclusions, Gonzalez Delgado \& Cid Fernandes (2009) commented that 
the STELIB-based models produced metallicities of the star clusters
systematically smaller by about 0.6\,dex with respect to that 
was found with other models, but ages were probably right.
The fact was that STELIB significantly underestimated metallicities  
with respect to both the CaII triplet and the metallicities from Leonardi \& Rose (2003). 
A similar conclusion was derived by Koleva et al. (2008)
in their analysis of Galactic globular clusters. 

\subsection{The age sequence }
\label{sec.sequence}

Since all the six types of sample galaxies are studied on 
their stellar populations, we can further study whether there is
any age sequence among them. This could be 
an extended study of 
Chen et al. (2009), in which we have shown that 
there is an age sequence
from star-forming galaxies, through composite galaxies, Seyfert 2s to LINERs,
since the young populations decrease following this sequence.

In addition to the information about the dependences of spectral synthesis results 
on stellar evolution track, 
the numbers in Table~\ref{bc03.alltype.atxt} also shows us a 
possible age sequence among six different galaxy types. 
In our earlier work, we have suggested that there was an age sequence: 
from star-forming, composite galaxies, Seyfert 2s to LINERs. 
Here we confirm this results, and we add E+A and early-type galaxies. 
It shows that the young populations decrease from
 star-forming, composite 
galaxies, through E+A galaxies, Seyfert 2s, LINERs to early-type galaxies. 
And E+A galaxies are dominated by 
intermediate-age populations, and the intermediate-age populations in E+A galaxies are 
far more than others. 
While early-type galaxies are certainly the oldest one among all galaxy types.
 Therefore,
we conclude there is a possible age sequence from star-forming, composite 
galaxies, through E+A galaxies, Seyfert 2s, LINERs to early-type galaxies.
A similar conclusion was reached by Schawinski et al. 2007, who identified an evolutionary 
sequence from star formation via nuclear activity to quiescence, i. e. from star-forming via 
transition region (something like the composite galaxies in our sample) and Seyfert AGN and 
LINER to quiescence. 
This is also consistent with Cid Fernandes et al. (2009b) and Stasinska (2008).

\begin{table}
\caption{Stellar populations in each galaxy type by using six SSPs (group No.(1)) 
from BC03 (columns 3-4) and GALEV (columns 5-6) in different cases of tracks, 
and the results from  BaSeL3.1 of BC03 (column 7). 
PL represents power law, and other symbols are the same as 
Table~\ref{sk.6ssp.allmodels}. 
}
\centering
\label{bc03.alltype.atxt}
\begin{tabular}{l|c|c|c|c|c|c}
\hline
galaxy types & age& \multicolumn{2}{c|}{BC03}& \multicolumn{2}{c|}{GALEV} &\\
\cline{3-6}
            &     bin     & Pa94 & Pa00& Pa94& Geneva&BaSeL3.1 \\
\hline
star-forming& Y&35 & 31 & 47 & 41 & 49 \\
& I            &54 & 51 & 39 & 48 & 40 \\
& O            &11 & 18 & 14 & 11 & 11 \\
\hline
composite& Y& 31 & 29 & 42 & 39 & 40 \\
& I         & 53 & 50 & 39 & 47 & 42 \\
& O         & 16 & 21 & 19 & 14 & 17 \\
\hline
E+A& Y& 13 & 9 & 22 & 20 & 19 \\
& I    & 77 & 71 & 71 & 72 & 72 \\
& O    & 10 & 20 & 7 & 8 & 9 \\
\hline
Seyfert 2& Y& 6 & 10 & 2 & 0 & 0 \\
& I         & 32 & 23 & 34 & 30 & 29 \\
& O         & 35 & 42 & 26 & 31 & 29 \\
& PL        & 27 & 25 & 38 & 39 & 42 \\
\hline
LINER& Y& 0 & 3 & 0 & 0 & 0 \\
& I     & 24 & 16 & 26 & 22 & 20 \\
& O     & 58 & 65 & 50 & 58 & 54 \\
& PL    & 18 & 16 & 24 & 20 & 26 \\
\hline
early type& Y& 6 & 7 & 0 & 0 & 0 \\
& I          & 7 & 2 & 40 & 20 & 31 \\
& O          & 87 & 91 & 60 & 80 & 69 \\
\hline
\end{tabular}
\end{table}

\section{Conclusion}
\label{sec.summary}

We compare 6 popularly used EPS models 
(BC03, CB07, Ma05, GALEV, GRASIL, Vazdekis/Miles) in this work. 
We use the SSPs they provide to fit the full optical spectra 
of six representative types of galaxies (star-forming and 
composite galaxies, Seyfert 2s, LINERs, E+A and early-type galaxies), 
which are taken from SDSS.
In details, we investigate the dependences of stellar population synthesis 
on EPS models, age, metallicity, and stellar evolution track.
We also study the age sequence of these different types of galaxies. 

\begin{enumerate}
\item The spectral fits of various galaxies by using different EPS models are excellent 
except possibly problematic regions, such as the area around $H_{\beta}$ line in few cases.

\item We select 6 SSPs with fixed ages, metallicities, IMF and stellar evolution tracks  
from each EPS model to fit the spectra 
of star-forming and E+A galaxies. 
So that we can explore the dependences of stellar 
population synthesis on EPS models. 
We comment that it is a complex job to do stellar population analysis
on galaxies, different EPS models may result in quite different 
results (such as BC03 vs. Vazdekis/Miles). 

\item We fix the IMF, metallicity, stellar evolution
tracks, and change the selection of ages to construct different 
SSP groups from the 6 EPS models, and then 
fit the spectra of star-forming and E+A galaxies. 
Thus we can study the effect of age selections on stellar population synthesis. 
The results show that the stellar population synthesis does
depend on the selections of ages. 

\item We fix the IMF, age, stellar evolution tracks,
 and change the selection of metallicities to construct more 
different SSP groups from the 6 EPS models. 
Then we fit the spectra of star-forming and E+A galaxies, 
in which way we can study the dependences of stellar population synthesis on the selection of 
metallicities. The results show that the stellar population synthesis also depends on the 
selection of metallicities, but which is less important than ages. 
We notice that this less dependence on metallicity than age may be due to either 
the way we show our results or the classic age-metallicity degeneracy. 

\item We fix the age and metallicity and change the selection of stellar evolution 
tracks in BC03 and GALEV models respectively. Then we use them to 
fit the spectra of 6 types of 
galaxies to check the effect of stellar evolution track on the stellar population 
synthesis. The results show that stellar evolution track does not affect 
much on the stellar population synthesis.

\item We also compare the stellar population synthesis results among different types of 
galaxies, and suggest that there is a possible age sequence: 
the importance of young populations decreases from star-forming, composite, Seyfert 2, LINER 
to early-type galaxies, and E+A galaxies lie between composite galaxies and Seyfert 2s in 
most cases. 

\end{enumerate}

\begin{acknowledgements}
We thank our referee for the valuable comments and suggestions, which helped improve this 
work. 
We thank Stephane Charlot and Gustavo Bruzual 
for kindly sending us the new version of their CB07 model,
also thank Jingkun Zhao for doing de-resolution for our spectra. 
X. Y. Chen also thanks Yue Wu for helpful discussions on using ULySS. 
We thank the NSFC grant support under Nos. 10933001, 10973006, 10973015, 10673002, and the 
National Basic Research Program of China (973 Program) Nos.2007CB815404, 2007CB815406, 
and No.2006AA01A120 (863 project). 

Funding for the SDSS and SDSS-II has been provided by the Alfred P. Sloan Foundation, the 
Participating Institutions, the National Science Foundation, the U.S. Department of Energy, 
the National Aeronautics and Space Administration, the Japanese Monbukagakusho, the Max 
Planck Society, and the Higher Education Funding Council for England. 
The SDSS Web Site is http://www.sdss.org/
 
The SDSS is managed by the Astrophysical Research Consortium for the Participating 
Institutions. The Participating Institutions are the American Museum of Natural History, 
Astrophysical Institute Potsdam, University of Basel, University of Cambridge, Case Western 
Reserve University, University of Chicago, Drexel University, Fermilab, the Institute for 
Advanced Study, the Japan Participation Group, Johns Hopkins University, the Joint 
Institute for Nuclear Astrophysics, the Kavli Institute for Particle Astrophysics and 
Cosmology, the Korean Scientist Group, the Chinese Academy of Sciences (LAMOST), Los 
Alamos National Laboratory, the Max-Planck-Institute for Astronomy (MPIA), the 
Max-Planck-Institute for Astrophysics (MPA), New Mexico State University, Ohio State 
University, University of Pittsburgh, University of Portsmouth, Princeton University, 
the United States Naval Observatory, and the University of Washington.

\end{acknowledgements}


\begin{thebibliography}{b}
 \bibitem[1993]{alongi} Alongi, M., Bertelli, G., Bressan, A., et al. 1993, \aaps, 97, 851
 \bibitem[2003]{anders} Anders, P. \& Alvensleben, U. Fritze-v. 2003, \aap, 401, 1063
 \bibitem[2007]{asari} Asari, N. V., Cid Fernandes, R., Stasi\'{n}ska, G., et al. 2007, \mnras, 381, 263
\bibitem[2009]{asari} Asari, N. V., Stasi\'{n}ska, G., Cid Fernandes, R., et al. 2009, ASPC, 408, 176
\bibitem[1981]{baldwin} Baldwin, J. A., Phillips, M. M. \& Terlevich, R.  1981, \pasp, 93, 5
\bibitem[1994]{bertelli} Bertelli, G., Bressan, A., Chiosi, C., Fagotto, F. \& Nasi, E. 1994, \aap, 106 275
\bibitem[1988]{bica} Bica, E. 1988, \aap, 195, 76
\bibitem[2000]{boisson} Boisson, C., Joly, M., Moultaka, J., Pelat, D. \& Roos, M. S. 2000, \aap, 357, 850
\bibitem[1993]{bressan} Bressan, A., Fagotto, F., Bertelli \& G., Chiosi, C. 1993, \aaps, 100, 647
 \bibitem[2004]{brinchmann} Brinchmann, J., Charlot, S., White, S. D. M., et al. 2004,  \mnras, 351, 1151
 \bibitem[1983]{bruzual} Bruzual, A. G. 1983, \apj, 273, 105
 \bibitem[1993]{bruzual} Bruzual, A. G. \& Charlot, S. 1993, \apj, 405, 538
 \bibitem[2003]{bruzual} Bruzual, A. G. \& Charlot, S. 2003, \mnras, 344, 1000
 \bibitem[2005]{bruzual} Bruzual, A. G. 2005, [astro-ph/0701907]
 \bibitem[1994]{calzetti} Calzetti, D., Kinney, A. L. \& Storchi-Bergmann, T. 1994, \apj, 429, 582
\bibitem[2009]{carter} Carter, D., Smith, D. J. B., Percival, S. M., et al. 2009, \mnras, 397, 695
\bibitem[1997]{cassisi} Cassisi, S., Castellani, M. \& Castellani, V. 1997, \aap, 317, 108
 \bibitem[1997]{cassisi} Cassisi, S., Degl'Innocenti, S. \& Salaris, M. 1997, \mnras, 290, 515
 \bibitem[2000]{cassisi} Cassisi, S., Castellani, V., Ciarcielluti, P., Piotto, G. \& Zoccali, M. 2000, \mnras, 315, 679
 \bibitem[2001]{cenarro} Cenarro, A. J., Cardiel, N., Gorgas, J., et al. 2001, \mnras, 326, 959
 \bibitem[2003]{chabrier} Chabrier, G. 2003, \pasp, 115, 763
 \bibitem[2009]{charlot} Charlot, S. \& Bruzual, A. G. 2009, in preparation
 \bibitem[2008]{chen} Chen, X. Y., Hao, C. N. \& Wang, J. 2008, \cjaa, 8, 25
 \bibitem[2009]{chen} Chen, X. Y., Liang, Y. C., Hammer, F., Zhao, Y. H. \& Zhong, G. H. 2009, \aap, 495, 457
\bibitem[2001]{cid fernandes} Cid Fernandes, R., Sodr\'{e}, L., Schmitt, H. R. \& Le\~{a}o, J.R.S. 2001, \mnras, 325, 60
\bibitem[2004]{cid fernandes} Cid Fernandes, R., Gu, Q., Melnick, J. et al. 2004, \mnras, 355, 273
\bibitem[2005]{cid fernandes} Cid Fernandes, R., Mateus, A., Sodr\'{e}, L., Stasi\'{n}ska, G. \& Gomes, J. M. 2005, \mnras, 358, 363
 \bibitem[2007]{cid fernandes} Cid Fernandes, R., Asari, N. V., Sodr\'{e}, L., et al. 2007, \mnras, 375, 16
\bibitem[2009]{cid fernandes} Cid Fernandes, R. et al. 2009a, The Starburst-AGN Connection, ASP Conference Series, ed. Wang, W. M., Yang, Z. Q., Luo, Z. J., Chen, Z., 408, 122
\bibitem[2009]{cid fernandes} Cid Fernandes, R., Schoenell, W., Gomes, J. M., et al. 2009b, RMxAC, 35, 127
\bibitem[2009]{cid fernandes} Cid Fernandes, R. \& Gonz\'{a}lez, D. 2009, [astro-ph/0912.0410]
\bibitem[2009]{coelho} Coelho, P. 2009, Probing Stellar Populations out to the Distant Universe: CEFALU 2008, AIP Conference Proceedings, 1111, 67
 \bibitem[2009]{conroy} Conroy, C., Gunn, J. E. \& White, M. 2009a, \apj, 699, 486
 \bibitem[2009]{conroy} Conroy, C., White, M. \& Gunn, J. E. 2009b, [astro-ph/0904.0002]
 \bibitem[2009]{conroy} Conroy, C. \& Gunn, J. E. 2009c, [astro-ph/0911.3151]
\bibitem[2007]{cordier} Cordier, D., Pietrinferni, A., Cassisi, S., \& Salaris, M. 2007, \aj, 133, 468
\bibitem[1962]{eggen} Eggen, O. J., Lynden-Bell, D. \& Sandage, A. R., 1962, \apj, 136, 748
\bibitem[2008]{eminian} Eminian, C., Kauffmann, G., Charlot, S., et al. 2008, \mnras, 384, 930
\bibitem[1972]{faber} Faber, S. M. 1972, \aap, 20, 361
 \bibitem[1994]{fagotto} Fagotto, F., Bressan, A., Bertelli, G. \& Chiosi, C. 1994a, \aaps, 104, 365
 \bibitem[1994]{fagotto} Fagotto, F., Bressan, A., Bertelli, G. \& Chiosi, C. 1994b, \aaps, 105, 29
\bibitem[2009]{falkenberg} Falkenberg, M. A. \& Fritze, U. 2009a, [astro-ph/0901.1665]
\bibitem[2009]{falkenberg} Falkenberg, M. A., Kotulla, R. \& Fritze, U., 2009b, [astro-ph/0905.0909]
\bibitem[1997]{fioc} Fioc, M. \& Rocca-Volmerange, B. 1997, \aap, 326, 950
\bibitem[1996]{girardi} Girardi, L., Bressan, A., Chiosi, C., Bertelli, G. \& Nasi, E. 1996, \aaps, 117, 113
 \bibitem[2000]{girardi} Girardi, L., Bressan, A., Bertelli, G. \& Chiosi, C. 2000, \aaps, 141, 371
 \bibitem[2007]{goto} Goto, T. 2007, \mnras, 381, 187
\bibitem[2009]{gonzalez}  Gonz\'{a}lez Delgado, R. M. \& Cid Fernandes, R. 2009, [astro-ph/0912.0413]
\bibitem[2005]{hammer}Hammer, F., Flores, H., Elbaz, D., et al. 2005, \aap, 430, 115
\bibitem[2007]{hammer} Hammer, F., Puech, M., Chemin, L., Flores, H. \& Lehnert, M. D., 2007, \apj, 662, 322
\bibitem[2009]{hammer} Hammer, F., Flores, H., Puech, M., et al. 2009, \aap, 507, 1313
\bibitem[2006]{hao} Hao, C. N., Mao, S. D., Deng, Z. G., Xia, X. Y. \& Wu H. 2006, \mnras, 370, 1339
\bibitem[2004]{heavens} Heavens, A., Panter, B., Jimenez, R., Dunlop, J. S., 2004, \nat, 428, 625
\bibitem[2005]{hempel} Hempel, M., Geisler, D., Hoard, D. W. \& Harris, W. E. 2005, \aap, 439, 59
\bibitem[2009]{huang} Huang, S. \& Gu, Q. S. 2009, [astro-ph/0906.2055]
\bibitem[2005]{jehin} Jehin, E., Bagnulo, S., Melo, C., Ledoux, C. \& Cabanac, R., The UVES Paranal Observatory Project: a public library of high resolution stellar spectra, in IAU Symp., ed. Hill, V., Francois, P., Primas, F. 2005, 261
\bibitem[2004]{jimenez} Jimenez, R., MacDonald J., Dunlop, J. S., Padoan, P. \& Peacock, J. A. 2004, \mnras, 349, 240
\bibitem[2003]{kauffmann} Kauffmann, G., Heckman, T. M., Tremonti, C., et al. 2003, \mnras, 346, 1055
\bibitem[1983]{kennicutt} Kennicutt, R. C. 1983, \apj, 272, 54
 \bibitem[1998]{kennicutt} Kennicutt, R. C. 1998, \araa, 36, 189
\bibitem[2001]{kewley} Kewley, L. J., Dopita, M. A., Sutherland, R. S., Heisler, C. A. \& Tervena, J. 2001, \apj, 556, 121
\bibitem[2006]{kewley} Kewley, L. J., Groves, B., Kauffmann, G. \& Heckman, T. 2006, \mnras, 372, 961
\bibitem[2008]{koleva} Koleva, M., Prugniel, Ph., Ocvirk, P., Le Borgne, D. \& Soubiran, C. 2008, \mnras, 385, 1998
\bibitem[2009]{koleva} Koleva, M., De Rijcke, S., Prugniel, Ph., Zeilinger, W. W. \& Michielsen, D. 2009a, \mnras, 396, 2133
\bibitem[2009]{koleva} Koleva, M., Prugniel, Ph., Bouchard, A. \& Wu, Y., 2009b, \aap, 501, 1269
\bibitem[1978]{koski} Koski, A. T. 1978, \apj, 223, 56
 \bibitem[2009]{kotulla} Kotulla, R., Fritze, U., Weilbacher, P., et al. 2009, [astro-ph/0903.0378]
 \bibitem[2006]{kriek} Kriek, M., van Dokkum, P. G., Franx, M. et al. 2006, \apj, 645, 44
 \bibitem[2007]{kriek} Kriek, M., van Dokkum, P. G., Franx, M. et al. 2007, \apj, 669, 776
 \bibitem[2008]{kriek} Kriek, M., van Dokkum, P. G., Franx, M. et al. 2008, \apj, 677, 219
 \bibitem[2001]{kroupa} Kroupa, P. 2001, \mnras, 322, 231
 \bibitem[1992]{kurucz} Kurucz R. L. 1992, in The Stellar Populations of Galaxies, ed. Barbuy, B. \& Renzini, A. IAU Symp., 149, 225
 \bibitem[2003]{leborgne} Le Borgne, J.-F., Bruzual, G., Pell\'{o}, R., et al. 2003, \aap, 402, 433
\bibitem[2004]{leborgne} Le Borgne, D., Rocca-Volmerange, B., Prugniel, P., et al. 2004, \aap, 425, 881
\bibitem[2007]{lee} Lee, H.-C., Worthey, G., Trager, S. C. \& Faber, S. M. 2007, \apj, 664, 215
\bibitem[1995]{leitherer} Leitherer, C. \& Heckman, T. M. 1995, \apjs, 96, 9
 \bibitem[1997]{lejeune} Lejeune, Th., Cuisinier, F. \& Buser, R. 1997, \aaps, 125, 229
 \bibitem[1998]{lejeune} Lejeune, Th., Cuisinier, F. \& Buser, R. 1998, \aaps, 130, 65
\bibitem[2003]{leonardi} Leonardi, A. J. \& Rose, J. 2003, \apj, 126, 1811
\bibitem[2009]{longhetti} Longhetti, M. \& Saracco, P. 2009, \mnras, 394, 774
\bibitem[1998]{maraston} Maraston, C. 1998, \mnras, 300, 872
 \bibitem[2005]{maraston} Maraston, C. 2005, \mnras, 362, 799
 \bibitem[2006]{maraston} Maraston, C., Daddi, E., Renzini, A., et al. 2006, \apj, 652, 85
 \bibitem[2007]{marigo} Marigo, P. \& Girardi, L. 2007, \aap, 469, 239
\bibitem[2006]{mateus} Mateus, A., Sodr\'{e}, L., Cid Fernandes, R., et al. 2006, \mnras, 370, 721
\bibitem[2006]{mathis} Mathis, H., Charlot, S. \& Brinchmann, J., 2006, \mnras, 365, 385
\bibitem[1979]{miller} Miller, G. E. \& Scalo, J. M. 1979, \apjs, 41, 513
\bibitem[2009]{muzzin} Muzzin, A., Marchesini, D., van Dokkum, P. G. 2009, \apj, 701, 1839
\bibitem[2007]{panter} Panter, B., Jimenez, R., Heavens, A. F. \& Charlot, S. 2007, \mnras, 378, 1550
\bibitem[2008]{panter} Panter, B., Jimenez, R., Heavens, A. F. \& Charlot, S. 2008, \mnras, 391, 1117
\bibitem[2008]{pessev} Pessev, P., M., Goudfrooij, P., Puzia, T. H. \& Chandar, R. 2008, \mnras, 385, 1535
\bibitem[1998]{pickles} Pickles, A. J. 1998, \pasp, 110, 863
\bibitem[2004]{pietrinferni} Pietrinferni, A., Cassisi, S., Salaris, M. \& Castelli, F. 2004, \apj, 612, 168
\bibitem[2009]{pracy} Pracy, M. B., Kuntschner, H., Couch, W. J., et al., 2009, [astro-ph/0903.4719]
\bibitem[2007]{prugniel} Prugniel, P., Soubiran, C., Koleva, M. \& Le Borgne, D. 2007, [astro-ph/0703658]
\bibitem[2005]{raimondo} Raimondo, G., Brocato, E., Cantiello, M. \& Capaccioli, M. 2005 \aj, 130, 2625
\bibitem[1955]{salpeter} Salpeter, E. E. 1955, \apj, 121, 161
 \bibitem[2006]{sanchez} S\'{a}nchez-Bl\'{a}zquez, P., Peletier, R. F., Jim\'{e}nez-Vicente, J. et al. 2006, \mnras, 371, 703
\bibitem[1998]{scalo} Scalo, J. 1998, in The Stellar Initial Mass Function, 38th, Herstmonceux Conference, ed. Gilmore, G. \& Howell D., ASP Conf. Ser., 142, 201
\bibitem[1992]{schaller} Schaller, G., Schaerer, D., Meynet, G. \& Maeder, A. 1992, \aaps, 96, 269
\bibitem[2007]{schawinski} Schawinski, K., Thomas, D., Sarzi, M., et al. 2007, \mnras, 382, 1415
\bibitem[2002]{schulz} Schulz, J., Alvensleben, U. Fritze-v., M\"{o}ller, C. S. \& Fricke, K. J. 2002, \aap, 392, 1
\bibitem[1981]{shuder} Shuder, J. M. \& Osterbrock, D. E. 1981, \apj, 250, 55
\bibitem[1998]{silva} Silva, L., Granato, G. L., Bressan, A. \& Danese, L. 1998, \apj, 509, 103
\bibitem[2008]{stasinska} Stasi\'{n}ska, G., Asari, V. N., Cid Fernandes, R., et al. 2008, \mnras, 391, 29
\bibitem[2002]{stoughon} Stoughon, C., Lupton, R. H., Bernardi, M., et al. 2002, \aj, 123, 485
\bibitem[1978]{tinsley} Tinsley, B. M. 1978, \apj, 222, 14
 \bibitem[2004]{tremonti} Tremonti, C. A., Heckman, T. M., Kauffmann, G., et al. 2004, \apj, 613, 898
 \bibitem[2004]{valdes} Valdes, F., Gupta, R., Singh, H. P. \& Bell, D. J. 2004, \apjs, 152, 251
 \bibitem[1996]{vazdekis} Vazdekis, A., Casuso, E., Peletier, R. F. \& Beckman, J. E., et al. 1996, \apjs, 106, 307
 \bibitem[1999]{vazdekis} Vazdekis, A. \& Arimoto, N. 1999, \apj, 525, 144
 \bibitem[1999]{vazdekis} Vazdekis, A. 1999, \apj, 513, 224
 \bibitem[2003]{vazdekis} Vazdekis, A., Cenarro, A. J., Gorgas, J., Cardiel, N. \& Peletier, R. F., et al. 2003, \mnras, 340, 1317
\bibitem[2009]{vazdekis} Vazdekis, A. 2009, in preparation
\bibitem[2005]{vazquez} V\'{a}zquez, G. A. \& Leitherer, C. 2005, \apj, 621, 695
\bibitem[1987]{veilleux}  Veilleux, S. \& Osterbrock, D. E. 1987 \apjs, 63, 295
 \bibitem[2002]{westera} Westera, P., Lejeune, T., Buser, R., Cuisinier, F. \& Bruzual, G. 2002, \aap, 381, 524
 \bibitem[1984]{white} White, S. D. M. 1984, \apj, 286, 38
 \bibitem[1994]{worthey} Worthey, G. 1994, \apjs, 94, 687
\end{thebibliography}
 \end{document}